%% file: 0.tex
\newif\ifshortversion
\shortversionfalse

\ifshortversion
\documentclass[cleveref, autoref]{lipics-v2021}
\else
\documentclass[reqno]{amsart}
\usepackage[foot]{amsaddr}
\fi

\usepackage{breakurl}             
\usepackage{hyperref}
\hypersetup{
  colorlinks=true,
  citecolor=blue,
  linkcolor=blue,
}
\RequirePackage{cleveref}

\usepackage{proof-at-the-end}
\ifshortversion
\pgfkeys{/prAtEnd/global custom defaults/.style={
    proof at the end,
    restate,
    no link to proof,
  }}
\else
\pgfkeys{/prAtEnd/global custom defaults/.style={normal}    }
\fi

\usepackage{comment}

\usepackage{tikz-cd}
\usepackage{ebproof}
\graphicspath{{../latex/figs/}}

\ifshortversion
\input{svFirstPage}
\fi

\input{macrosA}
\input{macrosB}

\input{theorems}

\newcommand{\END}{\end{document}}

\makeatletter
\def\input@path{{sections/}{./}}
\makeatother

\begin{document}
\ifshortversion
\else
\input{amsFirstPage}

\fi

\input{intro}

\input{preliminaries}

\input{dual}

\input{adjoint_maps}

\input{actions}

\input{frob}

\input{adjoints}

\input{fsttheo}

\input{sndtheo}

\input{conclusions}

\ifshortversion
\else
\bibliographystyle{plainurl}
\fi
\bibliography{biblio_FQSAC.bib}

\end{document}

%% file: macrosB.tex
\renewcommand{\SA}{\Star{A}}
\newcommand{\Fs}{Frobenius structure\xspace}

\newcommand{\parrows}[2]{\ar[shift left=1]{r}{#1} \ar[swap, shift right=1]{r}{#2}}

\newcommand{\proofofitem}[1]{\noindent Proof of \ref{#1}.~}

\renewcommand{\to}{\rto}
\newcommand{\sV}{\mathsf{V}}
\newcommand{\autonomous}{autonomous\xspace}
\newcommand{\aut}{\autonomous}
\newcommand{\ecat}{environment category\xspace}

\newcommand{\mate}{mate\xspace}

\newcommand{\ldcat}{linearly distributive category\xspace}

\ifshortversion
\else
\newcommand{\MyParagraph}[1]{\noindent{{\bf #1}}~}
\newcommand{\myVspace}[1]{}
\newcommand{\twoCols}[3][0.5]{
  \begin{center}
    #3
  \end{center}
  #2
}
\renewenvironment{wrapfigure}[3][]{\begin{center}}{\end{center}}
\newcommand{\ontheright}{}
\newcommand{\mySkip}{\medskip}
\newcommand{\myHfill}{}
\fi

%% file: abstract.tex
It is known that the quantale of sup-preserving maps from a complete
lattice to itself is a Frobenius quantale if and only if the lattice
is completely distributive. Since completely distributive lattices are
the nuclear objects in the autonomous category of complete lattices
and sup-preserving maps, we study the above statement in a categorical
setting. We introduce the notion of Frobenius structure in an
arbitrary autonomous category, generalizing that of Frobenius
quantale. We prove that the monoid of endomorphisms of a nuclear
object has a Frobenius structure. If the environment category is
star-autonomous and has epi-mono factorizations, a variant of this
theorem allows to develop an abstract phase semantics and to
generalise the previous statement. Conversely, we argue that, in a
star-autonomous category where the monoidal unit is a dualizing
object, if the monoid of endomorphisms of an object has a Frobenius
structure and the monoidal unit embeds into this object as a retract,
then the object is nuclear.

%% file: amsFirstPage.tex
\makeatletter
\newcommand{\definetitlefootnote}[1]{
  \newcommand\addtitlefootnote{
    \makebox[0pt][l]{$^{*}$}
    \footnote{\protect\@titlefootnotetext}
  }
  \newcommand\@titlefootnotetext{\spaceskip=\z@skip $^{*}$#1}
}
\makeatother

\definetitlefootnote{
  Work supported by the ANR project LAMBDACOMB ANR-21-CE48-0017}

\title[Frobenius structures]{Frobenius structures \\in star-autonomous categories\addtitlefootnote}
\author[C. De Lacroix]{Cédric De Lacroix}
\email{cedric.delacroix@lis-lab.fr}
\author[L. Santocanale]{Luigi Santocanale}
\email{luigi.santocanale@lis-lab.fr}
\address{
  LIS, CNRS UMR 7020, Aix-Marseille Universit\'e,
  France
}

\maketitle
\begin{abstract}
  \input{abstract.tex}
\end{abstract}


%% file: intro.tex
\section{Introduction}

\MyParagraph{Context.}  A major motivation for giving birth to linear
logic was to restore a classical negation in a constructive
setting. This was achieved on the proof-theoretic side, by controlling
the structural rules of weakening and contraction. As a byproduct, the
logical connectives of conjunction and disjunction have been split
into their multiplicative and additive versions, and the additive
connectives no longer have a classical behaviour, that is, they 
no longer distribute over each other.

We might tackle the same problem---restoring a classical negation in a
constructive setting---using algebraic and model theoretic approaches
and considering variants of linear logic.  A standard complete
provability semantics of (possibly non-commutative) linear logic is on
the class of structures known as quantales, see \eg
\cite{Yetter1990}. A quantale $(Q,\qmult)$ is a complete lattice with
a multiplication which distributes over sups in both variables (\ie,
it a semigroup in the category \SLatt of complete lattices and
join-preserving maps). Restoring classical negation means, in this
context, considering Frobenius or Girard quantales. A quantale is
Frobenius if it also comes with two suitable antitone maps
$\Lneg{\fun}$ and $\Rneg{\fun}$ representing negations, and it is
Girard if these two maps coincide.
Among quantales, probably the most interesting are those on the
set $\HLL$ of \jp endomaps of a complete lattice $L$, with
multiplication given by composition and the ordering computed
pointwise. These quantales naturally arise in logic, computer science,
and quantum mechanics.  For example, for a set $X$ and its powerset
lattice $P(X)$, the quantale $\homm{P(X)}{P(X)}$ is isomorphic to the
quantale $P(X\times X)$ of relations on $X$ and it is the main object
of study in relation algebra \cite{Maddux2006}.  Use of these
quantales to model concurrency and quantum mechanics/computation
abound in the literature, see \eg
\cite{AbramskyVickers1993,ACS1998,Resende2001}.
Our interest for quantales of the form $\HLL$ stems from generalized
linear orders
\cite{SAN-2018-RAMICS,SAN-2019-WORDS,SAN-2021-JPAA,S_RAMICS2021},
which are sort of fuzzy relations whose set of truth values is a
quantale. These relations need to satisfy contraints which are expressed
using the negation, hence the quantale must be a Frobenius quantale.
While developing a theory of these structures we came across (and
contributed to) the following result:
\begin{theorem}[See
  \cite{KrumlPaseka2008,EggerKruml2010,EGHK2018,S_RAMICS2020,S_ACT2020}]
  \label{theo:motivation}
  The quantale $\HLL$ of sup-preserving endomaps of a complete lattice
  $L$ is a 
  Frobenius quantale if and only if $L$ is completely distributive.
\end{theorem}
Completely distributivity is an infinitary generalization of the usual
distributive law. Basic examples of \cd lattices are the complete
linear orders, the powerset lattices and, of course, the finite
distributive lattices.
As logicians, the above theorem struck us, since it shows that a model
theoretic approach to restoring a classical negation fundamentally
diverges from the proof-theoretic one.  Indeed, the theorem has the
following reading: \emph{if a classical negation is enforced on the
  most standard class of models of non-commutative intuitionistic
  linear logic \textnormal{(the \jp endomaps of a complete lattice)},
  then also the additive connectives of the logic \textnormal{(the
    suprema and infima of the lattice)} have a classical behaviour
  \textnormal{(that is, they distribute over each other)}}.

\medskip

\MyParagraph{Contribution.} 
Due to its strength, the above statement deserves to be studied from
the largest number of perspectives. We take here a categorical
approach and give a proof of the statement that relies on the \saut
structure of $\SLatt$, the category of complete lattices and \jp
maps. In doing so, we generalise the statement to \sautcats. 
This is possible due to a fundamental characterization of complete
distributivity in the categorical language, stating that \emph{the
  \cdlatt{s} are exactly the nuclear objects in \SLatt}, see
\cite{Raney60,HiggsRowe1989}.
We recall that an object $A$ of a \smcc
$\V = (V,I,\tensor,\alpha,\lambda,\rho,\IHom{-,-},\ev)$ is
\emph{nuclear} if the canonical map $\mix : \SATA \to \HAA$ is an
isomorphism, where $\SA$, the dual of $A$, is the internal hom
$\IHom{A,\zero}$ (see e.g. \cite{Rowe1988}).

To achieve our goal, we define Frobenius structures in a \smcc by
strictly mimicking the definition of (possibly unitless) Frobenius
quantales given in \cite{deLacroixSantocanale2022}.
Definitions of similar structures 
appear in the literature
\cite{kock_2003,Hyland2004,Street2004,Egger2010} and differ from our
mainly w.r.t. the \ecat and the presence of units.  We consider the
choice of an axiomatization only as a tool for our goal of
generalizing Theorem~\ref{theo:motivation} to a categorical
setting. We insist, however, that the theorem is about characterizing
when a canonical quantale on the tensor product $\SLTL$ of an object
$L$ in $\SLatt$ has a unit.
To define \Fs{s}, we use the notion of \emph{dual pairing} in a
monoidal category $\V$. This is a map
$\epsilon : A \tensor B \rto \zero$ with two given universal
properties.
It turns out that in a \sautcat such a dual pairing exists if and only
if $B$ is isomorphic to $\SA$, if and only if $A$ is isomorphic to
$\SB$.
This notion provides the framework by which to study objects that are
dual to each other only up to isomorphism: for example $(\SATA,\HAA)$
is part of a dual pairing in any \sautcat and, for any \cl $L$,
$(L,L^{op})$ is part of a dual pairing in \SLatt.
If $\epsilon : A \tensor B \rto \zero$ is a dual pairing and $A$ is a
semigroup, then $A$ acts on $B$ on the left and on the right. The left
and right actions, noted $\lact[A]$ and $\ract[A]$, correspond, in the
category \SLatt, to the two implications of a quantale. 
We define generalized
Frobenius quantales in arbitrary symmetric monoidal category as follows:
\begin{definition*}
  A \emph{\Fs} is a tuple $\frob$ where $(A,B,\epsilon)$ is a dual
  pairing, $(A,\mu_A)$ is a semigroup, 
$\leftA$ and $\rightA$ are 
invertible maps from $A$ to $B$
such that
\begin{align}
  \label{eqs:FrobStructure}
    \epsilon \circ (A \tensor \leftA) & = \epsilon \circ (A \tensor \rightA)
    \circ \sigma_{A,A}
    \qquad\Tand\qquad \lact \circ (A \tensor r) = \ract[A] \circ (l
    \tensor A)\,.
  \end{align}
\end{definition*}
Almost by definition, in $\SLatt$, a \Fs of the form
$(Q,Q^{op},\epsilon,\qmult,\Lneg{\fun},\Rneg{\fun})$ amounts to a
Frobenius quantale, as defined in \cite{deLacroixSantocanale2022},
that is a quantale $(Q,\qmult)$ coming with antitone negations
satisfying
\begin{align*}
  x \leq \Lneg{y} & \;\;\tiff\;\; y \leq \Rneg{x} 
  \qquad\Tand\qquad \Rneg{y}
  \rlimpl x = y \lrimpl \Lneg{x}\,.
\end{align*}
We prove then the following result, generalizing the direct
implication of Theorem~\ref{theo:motivation}.
\begin{thmIntro}
  \label{thmIntro:direct}
  If $A$ is nuclear, then there is a map $l$ such that
  $(\HAA,\Star{\HAA},\ev_{\HAA,I},\circ,l,l)$ is a Frobenius
  structure.
\end{thmIntro}
The proof of this theorem is almost straightforward from the
definition of \Fs.  Let us mention that Theorem~\ref{thmIntro:direct}
is strictly related to (and may be considered an instance of)
Corollary 3.3 in \cite{Street2004}.  The connection, however, depends
on the fact that adjoints in a \smcc, or dualisable objects, are
exactly the nuclear objects, as argued in Section~\ref{sec:adjoints}.
The statement can be further generalised: if $\mix$ is not invertible
but the underlying \sautcat has some nice factorization system, then
the image of $\mix$ is the support of a Frobenius structure.  In
\SLatt, this construction yields the Girard quantale of tight endomaps
of a \cl studied in \cite{deLacroixSantocanale2022}.  The
generalization is a consequence of a double negation construction that
we described in \cite{deLacroixSantocanale2022} for Frobenius
quantales, and that we generalise here to a categorical setting. The
statement sounds as follows:
\begin{thmIntro}
  \label{thmIntro:mainthm}
  Let $\V$ be a \saut category with an epi-mono factorization system.
  Let $(A,\mu_{A})$ be a semigroup in $\V$ and let
  $\epsilon : A \tensor B \rto I$ be a dual pairing.  If
  $f : A \rto B$ is a map such that
  $\epsilon \circ (A \tensor f) = \epsilon \circ (A \tensor
  f) \circ \sigma_{A,A}$, then the image of $f$ is the carrier of a
  \Fs.
\end{thmIntro}

We also demonstrate that the converse of Theorem~\ref{thmIntro:direct}
actually holds if we add another condition, which we identify now as a
key ingredient of the proof in \cite{S_ACT2020} of
Theorem~\ref{theo:motivation}.
We say that an objet $A$ of a monoidal category is \emph{\paffine} if
the tensor unit $I$ embeds into $A$ as a retract. 
For example, every \cl which is not a singleton
  is \paffine in
\SLatt.
We prove:
\begin{thmIntro}
  \label{thmIntro:converse}
  If $A$ is a \paffine object of a \sautcat and the canonical monoid
  $(\HAA,\circ)$ is part of a Frobenius structure, then $A$ is
  nuclear.
\end{thmIntro}

\MyParagraph{Related Work.}  
Besides the works that we already mentioned, either those on the
theory of quantales \cite{KrumlPaseka2008,EggerKruml2010,EGHK2018} or
those generalizing the notion of Frobenius quantale to some kind
monoidal setting \cite{kock_2003,Hyland2004,Street2004,Egger2010}, we
also wish to mention that the notion of nuclearity, originally
conceived for Banach spaces \cite{Grothendieck1955,Rowe1988}, has been
by now generalised in several directions \cite{ABP1999,BCS2000}. For
example, this notion is generalised in \cite{ABP1999} to the category
of Hilbert spaces, with motivations from quantum mechanics. While our
initial motivations for developing this research were of a purely
logical and order-theoretic nature, we could recognize, via the
formalization in a categorical language and the notion of nuclearity,
the similarity of our questionings with those arising in this line of
research.  In particular, we could construct in
\cite{deLacroixSantocanale2022} a unitless Frobenius quantale from
trace class operators (\ie, nuclear endomaps) of an infinite
dimensional Hilbert space. We noticed that the questioning about
unitless structures is pervasive in this line of research
\cite{AbramskyHeunen2012,GG2019,CCP2021}. 
We describe in the last section our first non-conclusive remarks on
the scope of our results within this family of monoidal categories. We
are convinced that connecting with existing research on nuclearity and
categorical quantum mechanics is a research direction worth to be
fully explored.

\bigskip

\MyParagraph{Structure of the paper.}  After introducing elementary
notions in Section~\ref{sec:preliminaries}, we study the notion of
dual pairing in Section~\ref{sec:dualpairs} and give an overview of
actions in dual pairs in Section~\ref{sec:actions}. This makes it
possible to introduce Frobenius structures in
Section~\ref{sec:frob}. We then state in Section~\ref{sec:adjoints}
the equivalence between the notion of adjoint and that of nuclear
object in autonomous categories. We finally prove
Theorems~\ref{thmIntro:direct} and \ref{thmIntro:mainthm} in
Section~\ref{sec:fsttheo}, and Theorem~\ref{thmIntro:converse} in
Section~\ref{sec:sndtheo}.  We add a discussion of the results
obtained and sketch future research in Section~\ref{sec:conclusions}.

%% file: preliminaries.tex
\section{Background}

\label{sec:preliminaries}
\renewcommand{\zero}{0}

Let us recall that a \emph{quantale} is a semigroup in the \sautcat $\SLatt$,
see \cite{EGHK2018}.  Otherwise said, it is a pair $(Q,\qmult)$ with
$Q$ a \cl and $\qmult$ a semigroup operation which distributes in each
place with suprema.
An essential feature of a quantale $(Q,\qmult)$ are the two
implication maps $(-\lrimpl -):Q\tensor \Qd\rto \Qd$ and
$(-\rlimpl -):\Qd\tensor Q\rto \Qd$ satisfying the adjointness
relations
\begin{align*}
  x \qmult y & \leq z \Tiff x \leq z \rlimpl y  \Tiff y\leq x \lrimpl z\,.
\end{align*}
We rely on the standard monograph \cite{maclane} for elementary facts
and notation concerning categories.  We shall say that a category is
\emph{\autonomous} if it is symmetric monoidal closed. Thus, an
\autonomous category is of the form
$\V = (\sV,I,\tensor,\alpha,\lambda,\rho,\sigma,\IHom{-,-},\ev)$,
where $\sV$ is a category, $\tensor:\sV\times\sV\to \sV$ is a
bifunctor, $\homm{-}{Z} : \sV^{\op}\to \sV$ is the right adjoint to
$(X\tensor -)$, $\ev$ is the counit of this adjunction and
$\alpha_{X,Y,Z}:(X\tensor Y)\tensor Z\cong X\tensor (Y\tensor Z)$,
$\rho_X: X\tensor I\cong X$, $\lambda_X:I\tensor X\cong X$,
$\sigma_{X,Y}:X\tensor Y\cong Y\tensor X$ are all natural isomorphisms
satisfying well-known constraints. We shall work with strict monoidal
categories $\V$. If $\V$ is \autonomous and $\zero$ is a fixed object,
then we get a contravariant functor $\Star{\fun} \eqdef
\homm{-}{\zero}$. We define the map $j_X: X\to \SSX$ as the transpose
of $\ev_{X,\zero}\circ \sigma_{\SX,X}$.
 We say an object is \emph{reflexive} if $j_X$ is
an isomorphism. If every object is reflexive, then $0$ is said to be
dualizing and $\V$ is said to be
a \emph{\sautcat}.
 If $\zero =I$ then, we can define the morphism
$\mix_{X,Y}: \SYTX \to\HYX$ as the transpose of
$\lambda_X\circ \ev_{Y,I} $. An object $X$ is \emph{nuclear} if
$\mix_{X,X}$ is invertible. A nuclear object is necessarily reflexive.

\begin{definition}
  A \emph{magma} in a monoidal
  category $\V$ is a pair
  $(A,\mu_{A})$ with 
  $\mu_{A} : A \tensor A \rto A$ a morphism in $\V$.  A
  \emph{bracketed magma} in $\V$ is a triple $(A,\mu_{A},\pi_{A})$
  with $(A,\mu_{A})$ a magma and $\pi_{A} : A \tensor A \rto I$.  A
  bracketed magma is \emph{associative} if $(A,\mu_{A})$ is a
  semigroup in $\V$ (that is, if $\mu_{A}$ is associative) and
  $\pi_{A}$ is associative, meaning that the following two horizontal
  arrows are equal:
  \begin{center}
    \myVspace{-10pt}
    \begin{tikzcd}
      A \tensor A \tensor A \ar[shift left=2mm]{r}{\mu_{A} \tensor A}
      \ar[shift right=2mm, swap]{r}{A \tensor \mu_{A}}& A \tensor A
      \ar{r}{\pi_{A}} & I \,.
    \end{tikzcd}
  \end{center}
  \myVspace{-4pt}
  If $\V$ is symmetric monoidal and $(A,\mu_A)$ is a magma, a pairing
  $\pi_A$ is said to be \emph{co-associative} if
  $(A,\mu_{A}\circ \sigma_{A,A}, \pi_A)$ is an \abm.
\end{definition}

\begin{theoremEnd}
  [category=preliminaries]
  {fact}
  \label{lemma:associativeiffcoassoc}
  A pairing $\pi_A$ is associative if and only if the pairing $\pi_A\circ \sigma_{A,A}$ is co-associative.
\end{theoremEnd}
\begin{proofEnd}
  We show that if $\pi_A$ is associative, then $\pi_A\circ \sigma_{A,A}$ is co-associative.
  \begin{align*}
    \pi_{A} \circ \sigma_{A,A} \circ ((\mu_{A} \circ \sigma_{A,A})
    \tensor A) & = \pi_{A} \circ (A \tensor (\mu_{A}
    \circ \sigma_{A,A})) \circ \sigma_{A \tensor A, A} \\
    & = \pi_{A} \circ (A \tensor \mu_{A}) 
    \circ (A \tensor  \sigma_{A,A}) \circ \sigma_{A \tensor A, A} \\
    \pi_{A} \circ \sigma_{A,A} \circ (A \tensor (\mu_{A} \circ
    \sigma_{A,A}))
    & = \pi_{A} \circ ((\mu_{A}
    \circ \sigma_{A,A})\tensor A ) \circ \sigma_{A,A \tensor A} \\
    & = \pi_{A} \circ (\mu_{A} \tensor A) 
    \circ (\sigma_{A,A} \tensor A ) \circ \sigma_{A, A \tensor A}\,,
  \end{align*}
  thus the statement follows from the equality
  \begin{align*}
    (A \tensor  \sigma_{A,A}) \circ \sigma_{A \tensor A, A}
    & = (\sigma_{A,A} \tensor A ) \circ \sigma_{A, A \tensor A}
  \end{align*}
  and associativity of $\pi_{A}$.
\end{proofEnd}
Notice that if $(A,\mu_A)$ is a semigroup and
$\operatorname{tr} :A\rto I$ is any arrow, then
$\operatorname{tr}\circ \mu_A : A \tensor A \rto I$ is associative.

\begin{example}
  \label{example:SATAHAA_semigroup}
  In every \autcat, $(\HAA,\circ)$ is a well-known semigroup where
  $\circ$ is the internal composition defined as the transpose of
  $\ev_{A,A}\circ(\ev_{A,A}\tensor \HAA)$. Also, if $\zero = I$, then
  $(\SATA,\mu_{\SATA}, \pi_{\SATA} )$ is an associative bracketed
  magma with
  $\mu_{\SATA} \eqdef (\SA \tensor \lambda_{A})\circ (\SA \tensor
  \ev_{A,I} \tensor A)$ and
  $\pi_{\SATA}\eqdef \ev_{A,I}\circ\sigma_{\SA,A}\circ\mu_{\SATA}$,
  see e.g. \cite{KrumlPaseka2008}.
\end{example}

\begin{definition}
  \label{def:actions}
  Let $(A, \mu_A)$ be magma. A map $\alpha^{\ell}: A\tensor X \rto X$ is
  a \emph{left action} if the two parallel arrows
  \begin{center}
    \myVspace{-10pt}
    \begin{tikzcd}[column sep=15mm]
      A\tensor A \tensor X \ar[shift left=2mm]{r}{\mu_{A} \tensor X}
      \ar[shift right=2mm, swap]{r}{A \tensor \alpha^{\ell}}& A \tensor X
      \ar{r}{\alpha^{\ell}}& X
    \end{tikzcd}
  \end{center}
  \myVspace{-4pt}
  are equal. A map $\alpha^{\rho} : X\tensor A \rto X$ is defined to
  be a \emph{right action} in a similar way.
\end{definition}

\begin{definition}
  Let $(A,\mu_{A})$, $(B,\mu_{B})$ be two magmas.  A \emph{magma
    homomorphism} from $(A,\mu_{A})$ to $(B,\mu_{B})$
  is a map $f : A \rto B$ making the diagram below on the left
  commutative.  If $(A,\mu_{A},\pi_{A})$ and $(B,\mu_{B},\pi_{B})$ are
  bracketed magmas, a \emph{bracketed magma homomorphism} from
  $(A,\mu_{A},\pi_{A})$ to $(B,\mu_{B},\pi_{B})$ is a magma
  homomorphism $f : (A,\mu_{A}) \rto (B,\mu_{B})$ making the 
  diagram below in the middle commutative.
  \begin{center}
    \myVspace{-5pt}
    \begin{tikzcd}
      A \tensor A \ar{d}{\mu_{A}} \ar{r}{f \tensor f} & B \tensor B
      \ar{d}{\mu_{B}} \\
      A \ar{r}{f}& B
    \end{tikzcd}
    \qquad\qquad
    \begin{tikzcd}
      A \tensor A \ar{rd}{\pi_{A}} \ar{r}{f \tensor f} & B \tensor B
      \ar{d}{\pi_{B}} \\
      & I
    \end{tikzcd}
    \qquad\qquad
    \begin{tikzcd}
      A \ar{d}{f}\ar{r}{\tr{\pi_{A}}} & \SA \\
      B \ar{r}{\tr{\pi_{B}}} & \SB \ar{u}{\Star{f}}
    \end{tikzcd}
  \end{center}
\end{definition}
Let us notice that, when the \ecat is \autonomous, the diagram above
in the middle commutes if and only if the diagram on the right does.
\begin{theoremEnd}
  [category=preliminaries]
  {fact}
  \label{lemma:mix_is_semigroup_homomorphism}
  The arrow $\mix:\SATA\rto\HAA$ is a semigroup homomorphism.
\end{theoremEnd}
  \begin{proofEnd}
   We must check the commutativity of
   \begin{center}
    \includegraphics[page=1]{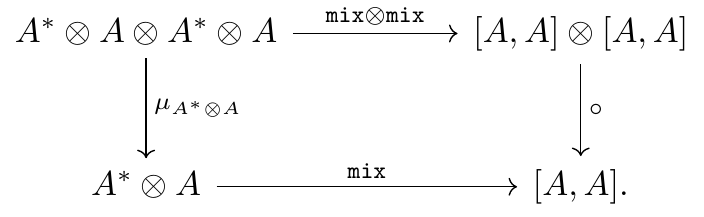}
  \end{center}
  Transposing and developing this diagram, we obtain
   \begin{center}
    \includegraphics[page=3]{mix_homomorphism.pdf}
   \end{center}
   which reduces to the commutativity of
   \begin{center}
    \includegraphics[page=4]{mix_homomorphism.pdf}
   \end{center}
   \end{proofEnd}

\begin{theoremEnd}
  {lemma}
  \label{prop:quotientSemigroup}
  Let $\V$ be \autonomous and let
  $e : (A,\mu_{A},\pi_{A}) \rto (B,\mu_{B},\pi_{B})$ be an epimorphism
  of bracketed magmas. If $(A,\mu_{A},\pi_{A})$ is associative, then
  so is $(B,\mu_{B},\pi_{B})$.
\end{theoremEnd}
\begin{proofEnd}
  Let us recall that since $\V$ is biclosed, then $X \tensor e$
  (resp. $e \tensor X$) is epi if $e$ is epi. Fron this, it follows
  that an arrow which is a tensor of epic arrows is epi.  We argue
  below that $\pi_{B}$ is associative.
  \begin{center}
    \begin{tikzcd}[column sep=13mm, ampersand replacement=\&]
      B \tensor B \tensor B \ar[swap]{d}{\mu_{B} \tensor B} \& A
      \tensor A \tensor A \ar[swap]{l}{e \tensor e \tensor e} \ar{r}{e
        \tensor e \tensor e} \ar[shift right=4mm, swap]{d}{\mu_{A} \tensor
        A} \ar[shift left=4mm]{d}{A \tensor \mu_{A}} \& B \tensor B
      \tensor B \ar[]{d}{B \tensor \mu_{B}} \\
      B \tensor B \ar[swap]{rd}{\pi_{B}} \& A \tensor A \ar[swap]{l}{e \tensor e } \ar{r}{e
        \tensor e } \ar{d}{\pi_{A}} \& B \tensor B \ar{ld}{\pi_{B}} \\
      \& I
    \end{tikzcd}
  \end{center}
  Since the two vertical parallel arrows from $A \tensor A \tensor A$
  are equal and all the squares and triangles commute, the left leg of
  the above diagram equals its right leg. Since
  $e \tensor e \tensor e$ is epi, we deduce that
  $\pi_{B} \circ (\mu_{B} \tensor B) = \pi_{B} \circ (B \tensor
  \mu_{B})$.
  
  The reader will immediatey adapt this argument to show that
  $\mu_{B}$ is associative.
\end{proofEnd}

%% file: dual.tex
\section{Dual pairings}
\label{sec:dualpairs}

\renewcommand{\zero}{0}

The goal of this section is to introduce the technical notion of dual
pairing, needed later to define \Fs{s}. We also present a few
properties of dual pairings. 
Throughout the section we let $\zero$ be a fixed object of an
environment monoidal category \V.
\begin{definition}
  A map
  $\epsilon : A \tesnor B \rto \zero$ in \V\ is said to be a
  \emph{dual pairing} (w.r.t. the object $\zero$) if the induced
  natural transformations
  \begin{align*}
    & \hom(X,B) \rto \hom(A\tensor X, \zero)\,, & \hom(X,A) \rto
    \hom(X\tensor B, \zero)\,,
  \end{align*}
  are isomorphims.
\end{definition}
Clearly, the above definition is equivalent to requiring that
$\epsilon$ has two universal properties: 1. \emph{for each
  $f : A\tensor X \rto \zero $ there exists a unique map
  $\tr{f}: X \rto B$ such that
  $\epsilon \circ (A \tensor \tr{f}) = f$}, and 2. \emph{for each
  $g : X\tensor B \rto \zero $ there exists a unique map
  $\tr{g}: X \rto A$ such that
  $\epsilon \circ (\tr{g} \tensor B)= g$}.  We call $\tr{f}$ (and
$\tr{g}$) the \emph{transpose} of $f$ (resp. $g$). For $f : X \rto B$
(resp. $g : X \rto A$), we let
$\trdown{f} = \epsilon \circ (A \tensor f)$ (resp.
$\trdown{g} = \epsilon \circ (g \tensor B)$).
If
$\epsilon: A \tensor B \rto \zero$ is a dual pairing, then we
shall also say that the triple $(A,B,\epsilon)$ is a dual
pairing, thus emphasizing the typing. We shall also informally say
that $(A,B)$ is a \emph{dual pair}, leaving aside the arrow
$\epsilon$.  A dual pairing
$\epsilon : A \tensor B \rto \zero$ is unique up to unique
isomorphism, as stated next.
\begin{theoremEnd}
  [category=dualpairs]
  {proposition}
  \label{lemma:uniquenessDual}
  If $(A,B_{1},\epsilon_{A,B_{1}})$ is a dual pairing and
  $\psi : B_{0} \rto B_{1}$ is iso, then
  $(A,B_{0},\epsilon_{A,B_{1}} \circ (A \tensor \psi))$ is a dual
  pairing. Conversely, if $(A,B_{0},\epsilon_{A,B_{0}})$ and
  $(A,B_{1},\epsilon_{A,B_{1}})$ are two dual pairings, then there
  exists a unique iso $\psi : B_{0} \rto B_{1}$ such that
  $\epsilon_{A,B_{0}} = \epsilon_{A,B_{1}} \circ (A \tensor \psi)$.
\end{theoremEnd}
\begin{proofEnd}
  We have to check that
  $\epsilon_{A,B_0}\eqdef \epsilon_{A,B_{1}} \circ (A \tensor \psi)$
  is a dual pairing. Indeed, we have two natural isomorphisms
  \begin{align*}  
&  \hom(X,B_0) \rto[\hom(X,\psi)] \hom(X,B_1)\rto[\trdown{\fun}] \hom(A\tensor X, \zero)\,,
  \\
 & \hom(X,A) \rto[\trdown{\fun}](X\tensor B_1, \zero)\rto[\hom(X\tensor \psi, \zero)] \hom(X\tensor B_0, \zero)\,,
\end{align*}
sending both $\id_{B_{0}}$ and, respectively, $\id_{A}$ to
$\epsilon_{A,B_0}$.

  Conversely, if $(A,B_{0},\epsilon_{A,B_{0}})$ and
  $(A,B_{1},\epsilon_{A,B_{1}})$ are two dual pairings,we have by definition two natural isomorphisms
  \begin{align*}
    \hom(X,B_{0}) & \xrightarrow{\trdown{\fun}} \hom(A \tensor X, I) \xrightarrow{\tr{\fun}} \hom(X,B_{1})\,,
  \end{align*}
  whose composition, using the Yoneda Lemma, is induced by
  post-composition with an invertible map $\psi : B_{0} \rto
  B_{1}$. Taking the image of $\id_{B_{0}}$, we see that
  $\tr{\epsilon_{A,B_{0}}} = \psi$, that is, we have
  $\epsilon_{A,B_0}= \epsilon_{A,B_{1}} \circ (A \tensor \psi)$.
\end{proofEnd}
Obviously, similar statements also hold with respect to the object on the
left, for example: if $(A_{0},B,\epsilon_{A_{0},B})$ and
$(A_{1},B,\epsilon_{A_{1},B})$ are two dual pairings, then there
exists a unique iso $\chi : A_{0} \rto A_{1}$ such that
$\epsilon_{A_{0},B} = \epsilon_{A_{1},B} \circ (\chi \tensor B)$.
While our presentation of the notion of dual pairing is symmetry-free,
we shall work later within symmetric and \autonomous categories. This
imposes a few remarks.
\vspace{-1mm}
\begin{theoremEnd}
  [category=dualpairs]
  {lemma}
  \label{lemma:DualPairSymmetry}  
  If \V\ is symmetric and $(A,B, \epsilon)$ is a dual pairing,
  then so is $(B,A,\epsilon \circ \sigma_{B,A})$.
\end{theoremEnd}
\begin{proofEnd}
  Clearly we have natural isomorphisms
\begin{align*}
  \hom(X,A) & \rto \hom(X \tensor B,\zero)
  \rto[\hom(\sigma_{B,X},\zero)] \hom(B \tensor X,\zero)
  \,, \\
  \hom(X,B) & \rto \hom(A \tensor X ,\zero)
  \rto[\hom(\sigma_{X,A},\zero)] \hom( X \tensor A,\zero) \,, 
  \end{align*}
  with universal arrows equal to
  $\epsilon\circ\sigma_{B,A}: B \tensor A \rto \zero$.    
\end{proofEnd}
\vspace{-1mm}
For $\V$ an autonomous category and an arrow $f : X \rto \SY$ in $\V$, we let
$\Perp{f} : Y \rto \SX$ be the transpose of the map 
\begin{tikzcd}
  X \tensor Y \ar{r}{\sigma_{X,Y}} & Y \tensor X
  \ar{r}{Y \tensor f}
  & Y \tensor \SY \ar{r}{\ev_{Y,\zero}} & \zero 
\end{tikzcd}\,.

Notice that $\Perp{\fun} : \hom(X,\SY) \rto \hom(Y,\SX)$ is natural in
$X$ and $Y$. If $g = \Perp{f}$, then we say that $f$ and $g$ are
\emph{\mate{s}}.  Let us remark that the canonical arrow
$j_X : X\rto \SSX$ is the \mate of $\id_{\SX}$, from which the
following statement easily follows:
\begin{theoremEnd}
   [category=dualpairs]
  {lemma}
  \label{lemma:perpFromStar}
  For $f : X \rto \SY$, $\Perp{f} = \Star{f} \circ j_{Y}$.
\end{theoremEnd}
\begin{proofEnd}
  Indeed, we have
  $\Perp{(\Star{f} \circ j_{Y})} = \Perp{(\Star{f} \circ
    \Perp{\id_{\SY}}j_{Y})} = (\id_{\SY})\Perp{}\Perp{} \circ f =
  \id_{\SY} \circ f = f$.
\end{proofEnd}
\vspace{-1mm}
Mates allow to precisely characterize dual pairs in autonomous
categories, as follows.
\vspace{-1mm}
\begin{theoremEnd}
   [category=dualpairs]
  {proposition}
  \label{prop:charDualPair}  
  If $\V$ is autonomous, then
  \begin{enumerate}
  \item \label{item:dualABstars} $(A,B)$ is a dual pair if and only if
    there are invertible \mate{s} 
    $\phi : A \rightarrow \SB$ and $\psi : B \rightarrow \SA$.
  \item \label{item:dualOfDualPair}  If $(A,B)$ is a dual pair, then $(\SA,\SB)$ is also a dual pair.
  \item  \label{item:reflexive} If $(A,B)$ is a dual pair, then both $A$ and $B$ are reflexive
    objects of \V.
  \item \label{item:charactReflexiveOne} If $A$ is reflexive, then $(A,\SA,\ev_{A,\zero})$ is a dual
    pairing.
  \item \label{item:charactReflexiveTwo} $(A,B)$ is a dual pair if and only if $A$ is reflexive and
    there is some iso $\psi : B \rto \SA$.
  \end{enumerate}
\end{theoremEnd}
\begin{proofEnd}

  \ref{item:dualABstars}. If $(A,B)$ is a dual pair, then we have natural isomorphisms
  \begin{align*}
    \hom(X,B) & \rto \hom(A \tensor X, \zero) \rto \hom(X,\SA)\,, 
  \end{align*}
  whose composition is induced, by the Yoneda Lemma, by
  post-composition with an invertible map $\psi : B \rto \SA$. Taking
  the image of $\id_{B}$, we see that $\tr{(\epsilon)} = \psi$.
  Similalry, the chain of natural isomorphisms
  \begin{align*}
    \hom(X,A) & \rto \hom(X \tensor B, \zero) \rto \hom(B \tensor X, \zero) \rto \hom(X,\SB)
  \end{align*}
  yields an invertible map $\phi : A \rto \SB$ satisfying
  $\tr{(\epsilon \circ \sigma_{B,A})} = \phi$. We have therefore
  $\phi = \Perp{\psi}$.

  Conversely, given those isomorphisms $\phi$ and $\psi$, we obtain
  the two natural isomorphisms as follows:
  \begin{align*}
    \hom(X,B) & \rto \hom(X,\SA) \rto \hom(A
    \tensor X,0) \,,
    \\
    \hom(X,A) & \rto \hom(X,\SB) \rto \hom(B
    \tensor X,0) 
     \rto
    \hom(X \tensor B,0)
    \,.
  \end{align*}

  The two universal arrows are, respectively, $\trdown{\psi}$ and
  $\trdown{\phi} \circ \sigma_{A,B}$ which are equal since we suppose
  $\phi = \Perp{\psi}$.

  \ref{item:dualOfDualPair}. If $(A,B)$ is dual, then $(B,A)$ is dual, by Lemma~\ref{lemma:DualPairSymmetry}. By Proposition~\ref{lemma:uniquenessDual}, dual pairs are closed   under isos. Since
    $B \simeq \SA$ and $A \simeq \SB$, the pair $(\SA,\SB)$ is dual.

  \ref{item:reflexive}.   Let $\phi : A\to \SB$ and $\psi : B\to \SA$ be the two isomorphims stated in
  \ref{item:dualABstars}. 
  Using Lemma~\ref{lemma:perpFromStar}, we obtain $    \phi  = \Perp{\psi} = \Star{\psi} \circ j_{A}$.  Since $\psi$ is invertible, $\Star{\psi}$ is invertible by
  functoriality, therefore $j_{A}$ is also an isomorphism and $A$ is a reflexive object.  By Lemma~\ref{lemma:DualPairSymmetry}, $B$ is reflexive as
  well.

  \ref{item:charactReflexiveOne}. By definition, $j_A$ is the transpose of $\ev_{A,0}\circ \sigma_{\SA,A}$, that is, $j_A = \Perp{1_{\SA}}$. Therefore, if $j_A$ is invertible then $(A,\SA,ev_{A,0})$ is a dual pair.

\ref{item:charactReflexiveTwo}. The direct implication is induced By \ref{item:dualABstars} and \ref{item:reflexive}. If $A$ is reflexive, then $(A,\SA)$ is a dual pair and if $\psi :B\to \SA$ is invertible then by lemma \ref{lemma:uniquenessDual}, $(A,B)$ is also a dual pair.
\end{proofEnd}

We next give some examples where the notion of dual pair naturally arises.

\begin{example}
\label{ex:dualpairLLop}
  Proposition~\ref{prop:charDualPair}.\ref{item:charactReflexiveTwo}
  shows that in a \sautcat $\V$, where all the objects are reflexive,
  any isomorphism $B \rto \SA$ is enough to build a dual pair
  $(A,B)$. This is a key observation for the coherence theorem in
  \cite{CHS2006}. For example, in \SLatt, taking as object $\zero$
  the two-element Boolean algebra $\two \eqdef \{\bot,\top\}$ (which  is also the
  unit of the tensor),
  the canonical isomorphism 
  $L^{\op} \iso \SL$ yields the dual pair $(L, L^{\op})$. 
  The pairing $\epsilon: L\tensor \Ld \rto \two$
  is given by $\epsilon(x,y) = \bot$, if $x \leq y$, and
  $\epsilon(x,y) = \top$, otherwise.
\end{example}

\begin{example}
  \label{example:SATAHAA}
  For any object $L$ of a \sautcat $\V$, the pair
  $(\SL \tensor L,[L,L])$ is dual with pairing 
  given by $\epsilon \eqdef  \ev_{L,\zero} \circ  \sigma_{\SL,L}\circ (\SL \tensor \ev_{L,L})$.
  Indeed, every object of $\V$ is reflexive, so
  $(\SLTL, \Star{(\SLTL)}, \ev_{\SLTL, \zero})$ is a dual pairing by
  Proposition~\ref{prop:charDualPair}.\ref{item:charactReflexiveOne}.
  It is well known that in a \sautcat the transpose of
  $\ev_{L,\zero}\circ \sigma_{\SL,L}\circ (\SL \tensor \ev_{L,L})$ is
  an isomorphism $\psi :\HLL \rto \Star{(\SLTL)}$. Then, by
  Proposition~\ref{lemma:uniquenessDual},
  $\ev_{\SLTL, \zero}\circ (\SLTL \tensor \psi) = \ev_{L,\zero} \circ
  \sigma_{\SL,L}\circ (\SL \tensor \ev_{L,L})$ is a dual pairing.

  \begin{theoremEnd}
    [category=dualpairs,all end]
    {claim}
    In a \saut category,
    The map $\psi :\HLL \rto \Star{(\SLTL)}$, transpose of  $
    \ev_{L,\zero} \circ  \sigma_{\SL,L}\circ (\SL \tensor \ev_{L,L})$,
    is an isomorphism.
  \end{theoremEnd}
  \begin{proofEnd}
    Let us verify that such a $\psi$ is an isomorphism.  Recall the
    usual natural isomorphisms
    \begin{align*}
      \hom(X, [L,L]) & \iso \hom(X, [L,\SSL]) \iso \hom(L \tensor X,
      \SSL) 
      \iso \hom(\SL \tensor L \tensor X,\zero) \iso \hom(X,\Star{(\SL
        \tensor L)})\,.
    \end{align*}
    By the Yoneda Lemma, the isomorphism from $\hom(X, [L,L])$ to
    $\hom(X,\Star{(\SL \tensor L)})$ is induced by the image of the
    identity in $\hom([L,L], [L,L]) $ along these isomorphisms, which
    we compute next:
    $$
    \begin{array}{rll}
      & [L,j_{L}] &\in \hom([L,L], [L,\SSL]) \,,\\
      j_{L} \circ \ev_{L,L} & = \ev_{L,\SL} \circ (L \tensor [L,j_{L}])
      & \in \hom(L \tensor [L,L], \SL)\,, \\
      \epsilon & \eqdef \ev_{L,\zero} \circ \sigma_{\SL,L} \circ (\SL \tensor
      \ev_{L,L}) \\
      &\; \; = \ev_{\SL,\zero} \circ (\SL \tensor j_{L} ) \circ (\SL \tensor
      \ev_{L,L}) & \in \hom(\SL\tensor L \tensor [L,L],\zero) \,,\\
      \psi & \eqdef \tr{\epsilon}&\in \hom( [L,L],\Star{(\SL\tensor L)})\,.
    \end{array}
    $$      
  \end{proofEnd}

\end{example}

\begin{example}
  \label{example:adjoints}
  It is also possible to understand dual pairs as a generalization of
  the well-known notion of \emph{adjunction}. Recall that a tuple $(A,B,\epsilon, \eta)$ with
  $\epsilon : A \tensor B \rto I$ and $\eta : I \rto B \tensor A $
  is said to be an adjunction if the two diagrams below
  commute. \\[-8pt]
  \twoCols[0.3]{
    $A$ is said to be left adjoint to $B$, $B$ right adjoint to
    $A$, $\eta$ is the unit of the adjunction and $\epsilon$ is the
    counit of the adjunction.  In an \aut category, \myHfill if}{
    \begin{center}
      \raisebox{-0.5\height}{\includegraphics[page=2,
        scale=1]{dualpairsDiags}}
    \end{center}
  }\\[2pt]
  $(A,B,\eta,\epsilon)$ is an adjunction, then so is
  $(B,A,\sigma_{A,B} \circ \eta, \epsilon \circ \sigma_{B,A})$, and
  therefore we won't distinguish between left and right adjoints, we
  just say that $A$ (or $B$) is part of an adjunction.\footnote{This
    also to avoid the naming conflict with the notion of adjoint,
    presented below in this section. An object $A$ which is part of an
    adjunction is often called \emph{dualizable}.}  It is a standard
  exercise in monoidal categories to verify the following statement.
  \begin{theoremEnd}
    [category=dualpairs]
    {fact}
    If
    $(A,B,\eta,\epsilon)$ is an adjunction, then $(A,B,\epsilon)$ is a
    dual pairing.
  \end{theoremEnd}
  \begin{proofEnd}
    Then $(A,B,\epsilon)$ is a dual pairing (where the object
    $\zero$ is $I$). Indeed, given $f : X \tensor B \rto I$, we get
    a map
    $\tr{f} : X \rto A$ defined by
    \begin{align*}
      \tr{f} & \eqdef X \iso X \tensor I \xrightarrow{X \tesnor
      \eta} X \tensor B \tensor A \xrightarrow{f \tensor A} I
      \tensor A \iso A\,.
    \end{align*}
    Clearly, $\epsilon \circ (\tr{f} \tensor B) = f$ and if
    $\epsilon \circ (g \tensor B) = f$, then the commutative diagram
    \begin{center}
      \includegraphics[page=3]{dualpairsDiags}
    \end{center}
    shows that $g = \tr{f}$. Thus, we have that
    $\hom(X \tensor B,I) \iso
    \hom(X, A)$.
    Observe also
    that $\tr{\epsilon} = \id_{A}$,
    by the first of the two diagrams defining an adjunction.
    With a similar argument, we show that the transformation sending
    $f : A \tensor X \rto I$ to
    $ \rho_{B} \circ (B \tensor f) \circ (\eta \tensor X) \circ
    \lambda_{X}^{-1} : X \rto B$ is a natural isomorphism and that
    $\epsilon$ corresponds to $\id_{B}$ under the isomorphism.
  \end{proofEnd}
  
  Not all the dual pairs arise from adjunctions. This can be simply
  observed by looking at the \sautcat $\SLatt$, where an object is
  part of an adjunction if and only if it is a nuclear object, by
  the content of Section~\ref{sec:adjoints}, if and only it is \cd,
  by \cite{Raney60,HiggsRowe1989}. Yet, $(L,L^{op},\epsilon)$,
  described in Example~\ref{ex:dualpairLLop}, is always a dual
  pairing, even when $L$ is not \cd.

  \noindent
  
  In \cite{Cockett99introductionto}, the authors introduce the
  notion of \emph{linear adjunction}. Recall that a linearly
    distributive category, see e.g. \cite{CockettSeely1997}, is a
    category with distinct monoidal structures $(I,\tensor)$ and
    $(0,\cotensor)$ coming with natural maps
    $\kappa^{\ell} : X \tensor (Y \cotensor Z) \rto (X \tensor Y)
    \cotensor X$ and
    $\kappa^{\rho} : (X \cotensor Y) \cotensor Z \rto X \cotensor (Y
    \cotensor Z)$ satisfying some constraints. A linear adjunction is
    then a tuple $(A,B,\epsilon, \eta)$ with
    $\epsilon : A \tensor B \rto \zero$ and $\eta : I \rto B \cotensor
    A$ making the diagrams on the\\
    \twoCols[0.3]{
      right commute.  \\ A \sautcat is, canonically, a \ldcat when we let
      $X \cotensor Y \eqdef \Star{(\Star{Y} \otimes \Star{X})}$.  In
      view of Example~\ref{ex:dualpairLLop} and Theorem 1.2 in
      \cite{Egger2010}, \hfill if the
    }{
      \begin{center}
        \raisebox{-0.5\height} {\includegraphics[page=7,
          scale=1]{dualpairsDiags}}
      \end{center}
    }\\
     \ecat is \saut, then $(A,B,\epsilon)$ is
    a dual pairing if and only if we can find
  $\eta : I \rto A \cotensor B$ such that $(A,B,\eta,\epsilon)$ is a
  linear adjunction.

\end{example}

%% file: adjoint_maps.tex
\MyParagraph{More on mates, adjoints.}
In \SLatt, for every \jp map $f:L\to M$ between complete lattices,
there exists a unique \jp map $\rho(f): \Md \to \Ld$ such that, for
each $x \in L$ and $y \in M$, $f(x)\leq y \ifff x\leq \rho(f)(y)$. The
map $\rho(f)$ is called the (right) \emph{adjoint} of $f$. The
equivalences defining the adjoint can be expressed diagrammatically,
by means of the equation
$\epsilon_{M,\Md} \circ (f\tensor \Md)= \epsilon_{L,\Ld} \circ
(L\tensor \rho(f))$.This suggests that a similar notion can be defined
for dual parings in an arbitrary monoidal category.

\medskip

\begin{wrapfigure}[4]{r}{0.3\textwidth}
\centering $
\begin{prooftree}
  \hypo{A_0\rto A_1} \infer1{A_0 \tensor B_1\rto 0} \infer1{B_1\rto
    B_0}.
\end{prooftree}
$
\end{wrapfigure}
Let $ (A_0,B_0,\epsilon_0)$ and $(A_1,B_1,\epsilon_1)$ be two dual
pairings. For a morphism $f: A_0\to A_1$, we denote by
$\Tilde{f} :B_1\to B_0$ the arrow given by the two transposes on the
right.  We say that $f : A_0 \rto A_1$ is \emph{adjoint} to
$g:B_1\rto B_0$ if $\Tilde{f} = g$.

 \begin{wrapfigure}[6]{r}{0.3\textwidth}
   \centering $
   \begin{tikzcd}[ampersand replacement=\&]
     X_0\tensor Y_1 \ar[]{d}{X_0\tensor g} \ar[]{r}{f\tensor Y_1} \&
     Y_0\tensor Y_1 \ar[]{d}{p_Y}
     \\
     X_0\tensor X_1 \ar[]{r}{p_X} \& 0 \,.
   \end{tikzcd}
   $
 \end{wrapfigure}
 A dual pairing $(A,B,\epsilon)$ is a special object of the category
 $\chu$, see \cite{Barr1979}.  Recall that a morphism between two
 objects $(X_0,X_1, p_X)$ and $(Y_0,Y_1, p_Y)$ of $\chu$ is a pair of
 $\V$ morphisms $(f:X_0\to Y_0,g:Y_1\to X_1)$ such the square on the
 right commutes.
 If $ (A_0,B_0,\epsilon_0)$ and $(A_1,B_1,\epsilon_1)$ are dual
 pairings, then a $\chu$ morphism between them is necessarily of the
 form $(f,\Tilde{f})$ and, also, a pair $(f,\Tilde{f})$ is obviously a
 morphism in $\chu$.  Denoting by $\Pair$ the full subcategory of
 $\chu$ whose objects are dual pairings, these remarks amount to the
 following statement, whose most relevant consequences is presented
 immediately after.
\begin{lemma}
  The obvious projection functors, $\mathcal{P}_0 : \Pair \rto \V$ and
  $\mathcal{P}_1 : \Pair \rto \V^{\op}$, are full and faithful.
\end{lemma}
\begin{theoremEnd}
  [category=dualpairs]
  {corollary}
  \label{lemma:inverseftilde}
  Let $(A_i,B_i,\epsilon_i)$, $i = 0,1,2$, be dual pairings. If
  $f:A_0\to A_1$ and $g : A_{1} \rto A_{2}$, then
  $\Tilde{\id_{A_{i}}} = \id_{B_{i}}$ and
  $\Tilde{g}\circ \Tilde{f} = \Tilde{(f\circ g)}$.  In particular, if
  $f:A_0\to A_1$ is inverted by $g:A_1\to A_0$, then $\Tilde{f}$ is
  inverted by $\Tilde{g}$.
\end{theoremEnd}
\begin{proofEnd}
  The two equalities $\Tilde{\id_{A_{i}}} = \id_{B_{i}}$ and
  $\Tilde{g}\circ \Tilde{f} = \Tilde{(f\circ g)}$
  are direct consequence of the
  functoriality of $ \mathcal{P}_0$ and $\mathcal{P}_1$. We directly
  deduce the last statement by
  $\Tilde{f}\circ \Tilde{g} = \Tilde{(g\circ f)}=
  \Tilde{\id_{A_0}}=\id_{B_0}$.
\end{proofEnd}
Let $X,Y$ be reflexive, so $(X,\SX,\ev_{X,0})$, $(Y,\SY_{Y,0})$, and
$(\SY,Y,\ev_{Y,0} \circ \sigma_{\SY,Y})$ are dual pairings. Notice the
following relations: for $f : X \rto Y$, $\Tilde{f} = \Star{f} $, that
is, the adjoint of $f$ is the result of the functorial action on the
map $f$; for $g : X \rto \SY$, $\Tilde{g} = \Perp{g}$, that is, the
adjoint of $g$ is just its mate.  This yields a number of
relations. For example, if also $Z$ is reflexive, $g : Z \rto Y$ and
$f : X \rto \SY$, then 
using   Corollary~\ref{lemma:inverseftilde}
we have
\begin{align}
  \label{eq:naturalityPerp}
  \Perp{(\Star{g} \circ f)} & = \Perp{f} \circ g : Z \rto \SX\,.  
\end{align}
Notice that equation~\eqref{eq:naturalityPerp} amounts to the
naturality of $\Perp{\fun} : \hom(X,\SY) \rto \hom(Y,\SX)$.

Finally, let us mention that we shall focus on maps $f : A \rto B$, where
$(A,B,\epsilon)$ is a dual pairing and, consequently,
$(B,A, \epsilon \circ \sigma_{B,A})$ as well. The following statement
is then easily verified.
\begin{lemma}
  \label{lemma:transposes}
  For $(A,B,\epsilon)$ a dual pairing and $f : A \rto B$, the
  transposes of $f$ and $\Tilde{f}$ differ by a symmetry:
  $\epsilon \circ (A \tensor \Tilde{f})  =
    \epsilon \circ (A \tensor f) \circ \sigma_{A,A}$.
\end{lemma}

%% file: actions.tex
\section{Actions in dual pairs}
\label{sec:actions}

The implications of a quantale validate the two equations
\begin{align*}
  z\rlimpl (y\qmult x) =(z\rlimpl y)\rlimpl x && \text{and} &&(x\qmult y) \lrimpl z = x\lrimpl (y \lrimpl z)\,.
\end{align*}
These two equations can be understood by saying that the implications
are, respectively, left and right actions of $Q$ over $\Qd$, see
Example~\ref{ex:action_in_a_quantale}.
In this section we show that actions arise from dual pairs when one of the
objects of a dual pair is a semigroup. From now on, $\V$ will be a
symmetric monoidal category.
Let $(A, B, \epsilon)$ be a dual pairing in $\V$ such that $(A,\mu_A)$
is a magma. We define two morphisms
\begin{align*}
   \lact:A\tensor B\rto B &&\text{and}&& \ract[A] :B\tensor A \rto B\,,
\end{align*}
as the only morphisms making these two diagrams commute:
\begin{align}
  \label{eq:left_act}
  & \raisebox{-0.5\height}{\includegraphics[scale=1,page=1]{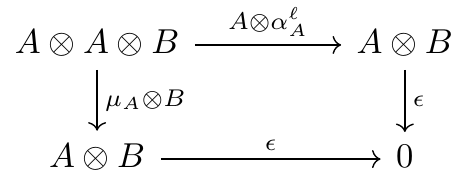}}
  & \raisebox{-0.5\height}{\includegraphics[scale=1,page=3]{defActs}}
\end{align}

\begin{theoremEnd}
  [category=actions]
  {fact}
  Letting $A^{op}$ be the magma
  $(A,\mu_{A}\circ \sigma_{A,A})$, the relation
  $\lact[A^{op}] = \ract[A] \circ \sigma_{A,B}$ holds.
\end{theoremEnd}
\begin{proofEnd}
  The fact is then an immediate cosequence.
  The diagram on the right of \eqref{eq:left_act} commutes
  if and only if the diagram below commutes:
  \begin{center}
    \includegraphics[scale=1,page=6]{defActs}
  \end{center}
  commutes. 
\end{proofEnd}

\begin{example} \label{ex:action_in_a_quantale}
As mentioned in Example~\ref{ex:dualpairLLop}, $(Q, \Qd)$ is a dual
pair in \SLatt. If $(Q,\qmult)$ is a quantale, commutativity of the
left (resp. right) diagram in \eqref{eq:left_act} amounts to the
relation
\begin{align*}
  x \qmult y & \leq z \Tiff x \leq \lact[Q](y,z)\qquad\qquad
  \text{(\,resp.,\quad$x \qmult y \leq z \Tiff y \leq \ract[Q](z,x)$\,)} \,.
\end{align*}
Therefore, by the uniqueness of the adjoint, we have
$\lact[Q](y,z) = z \rlimpl y$ and $\ract[Q](z,x) = x \lrimpl z$.
\end{example}

 \begin{example}
   Let $A$ be a finite dimensional $k$-algebra, $x,y \in A$,
   $l \in \SA= \homm{A}{k}$. It is direct to verify that
   $\lact(y, l):x\mapsto l(xy)$ and $\ract[A](l,x): y\mapsto l(xy)$.
 \end{example}

Let us give some elementary properties of $\lact$
and $\ract[A]$. 
\begin{theoremEnd}
  [category=actions]
  {lemma}
  \label{lemma:actions}
  Let $(A,B,\epsilon)$ be a dual pairing and $(A,\mu_A)$ be a magma.
  \begin{enumerate}
    \item \label{lemma:actions_item:relationleft_right}  
     The following equation, relating $\lact[A]$ and $\ract[A]$,
    holds:
    \begin{align}\label{flat_ract_and_flact_lactA}
      \epsilon \circ (A \tensor \ract[A]) & = \epsilon \circ
      \sigma_{B,A} \circ (\lact \tensor A)\,.
    \end{align}
  \item \label{lemma:actions_item:left_right_action} If $(A,\mu_{A})$
    is a semigroup, then the map $\lact$ is a left action (see
    Definition~\ref{def:actions}) and the map $\ract[A]$ is a right
    action.
  \item \label{lemma:actions_item:left_right_associativity} 
  If
    $(A,\mu_{A})$ is a semigroup, then the two actions are equivariant
    to each other, that is
\begin{align*}
  \lact\circ (A\tensor \ract[A]) = \ract[A] \circ (\lact \tensor A)\,.
\end{align*}

  \end{enumerate}
\end{theoremEnd}
\begin{proofEnd}
  \proofofitem{lemma:actions_item:relationleft_right}
  The equation is given by the commutativity of the diagram
  \begin{center}
    \includegraphics[page=1]{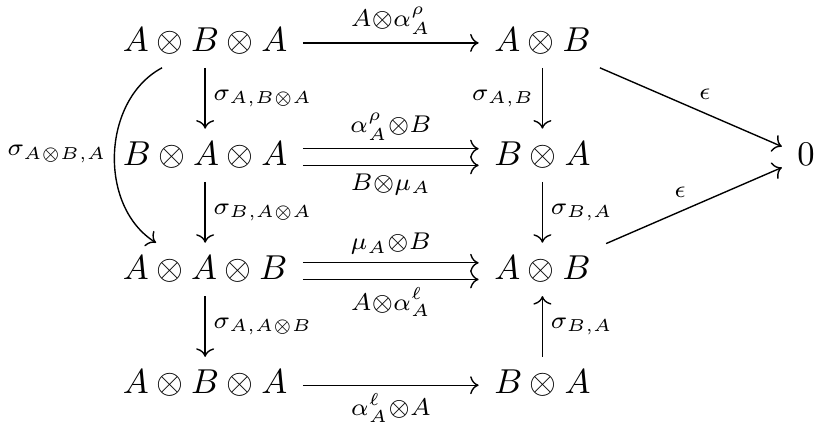}
  \end{center}
  
  \proofofitem{lemma:actions_item:left_right_action}
  In the following  diagram
  \begin{center}
    \begin{tikzcd}[ampersand replacement=\&]
      A^{3} \tensor B 
      \ar{dd}{A^{2} \tensor \lact} 
      \ar{rd}{\mu_{A} \tensor      A \tensor B} 
        \ar{rr}{A \tensor \mu_{A} \tensor B} 
        \&        \& A^{2} \tensor B
      \ar{d}{\mu_{A} \tensor B} 
      \ar{r}{A \tensor \lact} 
      \& A \tensor B
       \ar{dd}{\epsilon}
       \\
      \& A^{2} \tensor B 
      \ar{d}{A \tensor \lact}
      \ar{r}{\mu_{A} \tensor B} 
      \& A \tensor B \ar{rd}{\epsilon} 
      \& 
      \\
      A^{2} \tensor B 
      \ar{r}{\mu_{A} \tensor B} 
      \ar[swap]{rrd}{A \tensor \lact}
      \& A \tensor B 
      \ar{rr}{\epsilon} 
      \&\& 0 
      \\
      \&\& A \tensor B 
      \ar[swap]{ru}{\epsilon}
    \end{tikzcd}
  \end{center}
  the upper leg and lower leg are the transposes of the two parallel
  maps that are required to be equal for $\lact$ to be a left
  action.
  If $(A,\mu_{A})$ is a semigroup, then the above diagram is
  commutative, and so these two maps are indeed equal.

  The second
  statement follows since $(B,A,\epsilon)$ is a dual pair in the
  monoidal category $\V'$with same objects and arrows, where however
  we have $X \tensor' Y = Y \tensor X$.

   \proofofitem{lemma:actions_item:left_right_associativity}
  By \ref{lemma:actions_item:left_right_action}, $\ract[A]$ and $\lact[A]$ are two actions, therefore the diagram
  \begin{center}
    \includegraphics[page=2]{flat_ract_flat_lact}
  \end{center}
  commutes.
\end{proofEnd}

As the dual pair $(\HAA, \SATA)$ is the main  object studied in this paper, we characterize next the actions of $(\HAA,\circ)$ over
$\SATA$, 
and of $(\SATA,\mu_{\SATA})$ over $\HAA$,
see Examples~\ref{example:SATAHAA_semigroup} and
\ref{example:SATAHAA}.

\begin{theoremEnd}
  [category=actions]
  {proposition}
  \label{prop:action_SATA_et_HAA}
  In a \saut category $\V$, we have
  \begin{align*}
    \ract[\HAA] &= \SA \tensor \ev_{A,A}\,, &
    \lact[\HAA] & = \mu_{A,A,0} \tensor A\,.
  \end{align*}
\end{theoremEnd}
\begin{proofEnd}
  Clearly, the following diagram commutes:
  \begin{center}
  \includegraphics[page=1]{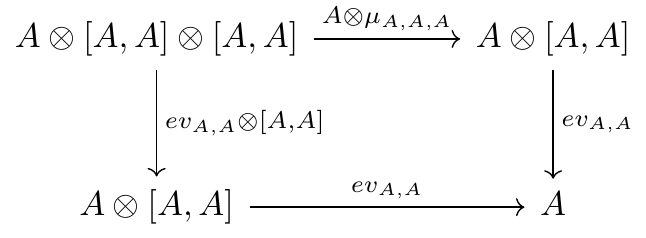}
  \end{center}
  By tensoring on the left with $\SA$ and then by postcomposing with
  $tr \eqdef \ev_{A,0} \circ \sigma_{\SA,A}$, we obtain commutativity of
  \begin{center}
  \includegraphics[page=2]{sndMainthmDiags}
  \end{center}
showing that $\ract[\HAA] = \SA \tensor \ev_{A,A}$.  In similar way,
the following diagram commutes
\begin{center}
  \includegraphics[page=3]{sndMainthmDiags}
\end{center}
and (by tensoring on the right with $A$) implies that
\begin{align}
  \label{diag:ract}
  \includegraphics[page=4]{sndMainthmDiags}
\end{align}
commutes, thus yielding  $\lact[\HAA] = \mu_{A,A,0} \tensor
A$. Indeed, we have
\begin{align*}
  \epsilon_{B,A} & = \epsilon_{A,B} \circ \sigma_{B,A} =
  \ev_{A,0}\circ \sigma_{\SA,A}\circ (\mu_{A,A,0} \tensor \SA)
\end{align*}
as witnessed by the diagram below:
\begin{align*}
  \includegraphics[page=5]{sndMainthmDiags}
  \tag*{\qedhere}
\end{align*}

\end{proofEnd}

%% file: frob.tex
\section{Frobenius structures}
\label{sec:frob}

We finally define in this section the notion of \Fs{} in a symmetric
monoidal category $\V$ and give some elementary properties of this
structure. While other approaches are possible, notably \cite{Street2004,Egger2010}, they are not strictly equivalent,
starting from the assumptions on the \ecat. \\[2pt]
\twoCols[0.65]{
  \begin{definition}
    \label{def:FrobStructure}
    A \emph{Frobenius structure} is a tuple $\frob$ where
    $(A,B,\epsilon)$ is a dual pairing, $(A,\mu_A)$ is a semigroup,
    and $\leftA$ and $\rightA$ are adjoint invertible maps from $A$
    to $B$ such that diagram~\eqref{diag:serre} \ontheright commutes.
  \end{definition}
}{
  \myVspace{-10pt}
  \begin{align}\label{diag:serre}
    \begin{tikzcd}[ampersand replacement=\&]
      A \tensor A \ar[swap]{d}{\leftA \tensor A} \ar{r}{A \tensor
        \rightA} \& A \tensor B
      \ar{d}{\lact[A]}\\
      B \tensor A \ar[swap]{r}{\ract[A]} \& B.
    \end{tikzcd}
  \end{align}
}
\myVspace{-0pt}
Notice that, by Lemma~\ref{lemma:transposes}, the above definition
coincides with the one given in the Introduction.

\myVspace{-5pt}
\begin{example}
  Definition~\ref{def:FrobStructure} strictly mimics the definition
  of Frobenius quantales in \cite{deLacroixSantocanale2022}.
  A Frobenius quantale is defined there as a tuple
  $(Q,\qmult,\Lneg{\fun},\Rneg{\fun})$ such that $(Q,\qmult)$ is a
  quantale, and where $(\Lneg{\fun},\Rneg{\fun})$ form an invertible
  Galois connection satisfying the equation  
    \begin{align}
      y \lrimpl \Lneg{x} & =  \Rneg{y} \rlimpl x\,.
      \label{eq:serre}
    \end{align}
    The commutative diagram~\eqref{diag:serre} is just a direct
    translation of equation~\eqref{eq:serre}, called the
    contraposition law in \cite{GJKO}.
    
    Let us recall a few considerations developed in
    \cite{deLacroixSantocanale2022}.  If a quantale has a dualizing
    element $0$, then the two maps $\Lneg{\fun}\eqdef 0\rlimpl \fun $
    and $\Rneg{\fun}\eqdef \fun \lrimpl 0 $ are inverse to each other,
    form a Galois connection, and satisfy equation
    \eqref{eq:serre}. And if a Frobenius quantale has a unit $1$ then
    $0 \eqdef \Lneg{1} = \Rneg{1}$ is a dualizing element. That is,
    contrarily to the usual definition 
    (see for instance
    \cite{Yetter1990,Rosenthal1990a,KrumlPaseka2008,EGHK2018}), a
    Frobenius quantale does not need a unit.\footnote{As a matter of
      fact, the theorems that we shall present in the next two
      sections can be understood as characterizing the existence of
      units in specific cases.}  Now it is immediate to see that
    Frobenius quantales as defined in \cite{deLacroixSantocanale2022}
    are exactly the Frobenius structures $(A,B,\epsilon,\mu,l,r)$ in
    \SLatt for which $B = A^{\op}$.
  \end{example}

Before giving other examples of \Fs{s}, let us give some characterization of these structures. 
In a quantale, equation~\eqref{eq:serre} is equivalent to the
shift/associative relations:
$$
x \qmult y \leq \Lneg{z}
\quad \Tiff\quad
\Rneg{x} \geq y \qmult z
\quad \Tiff\quad
x \leq \Lneg{(y \qmult z)}
\,.
$$
This also holds in the general setting, simply by transposing diagram
\eqref{diag:serre} we have:
\twoCols[0.47]{
  \begin{theoremEnd}
    [category=frob]
    {lemma}
    \label{lemma_shift_diagram}
    A pair of morphisms $(l,r)$ makes diagram~\eqref{diag:serre}
    commute if and only if the shift diagram~\eqref{diag:shift} \ontheright commutes.
  \end{theoremEnd}
}{
  \begin{align}
    \label{diag:shift}
    \raisebox{-0.45\height}{
      \includegraphics[page=1]{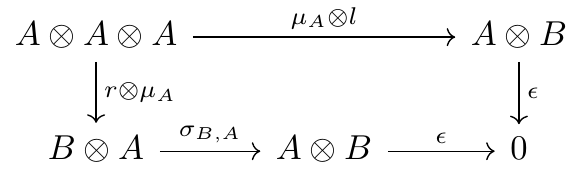}
    }
  \end{align}
}
\begin{proofEnd}
  Diagram~\eqref{diag:serre} commutes if and only if its transpose
  commutes:
  \begin{center}
    \includegraphics[page=2]{shift_diagram}
  \end{center}
  By \eqref{eq:left_act}, the upper leg above equals to $\epsilon
  \circ (\mu_{A} \tensor r)$:
  \begin{center}
    \includegraphics[page=3]{shift_diagram}
  \end{center}
  By equation~\eqref{flat_ract_and_flact_lactA}, the lower leg
  above equals to
  $\epsilon \circ (\mu_{A} \tensor l) \circ \sigma_{A \tensor A, A}$:
  \begin{center}
    \includegraphics[page=4]{shift_diagram}
  \end{center}
  Thus, diagram~\eqref{diag:serre} commutes if and only if
  \begin{align*}
    \epsilon
    \circ (\mu_{A} \tensor r) & = \epsilon \circ (\mu_{A} \tensor l) \circ \sigma_{A \tensor A, A}\,.
  \end{align*}
\end{proofEnd}

\vspace{-1mm}

Our next goal is to show that, for a \Fs $\frob$,  the dual pair
$(B,A)$ also carries a \Fs, in particular $B$ carries a canonical
semigroup structure.
\twoCols[0.62]{
\begin{theoremEnd}
  [category=frob]
  {lemma} 
  \label{lemma:Serre2}
  Let $(A,B,\epsilon)$ be a dual pair, $(A,\mu_A)$ a semigroup, and
  $(l,r)$ invertible adjoint maps from $A$ to $B$. The tuple $\frob$
  is a \Fs if and only if diagram~\eqref{diag:serre2} on the right
  commutes.
\end{theoremEnd}
}{
  \vspace{-10pt}
  \begin{align}
    \label{diag:serre2}
    \begin{tikzcd}[ampersand replacement=\&]
      B \tensor B 
      \ar[dotted]{dr}{\mu_{B}}
      \ar{r}{B \tensor \rightA^{-1}}
      \ar[swap]{d}{\leftA^{-1} \tensor
        B}
      \& B \tensor A
      \ar{d}{\ract[A]}
      \\
      A \tensor B
      \ar[swap]{r}{\lact[A]}
      \& B
    \end{tikzcd}
  \end{align}
}

\begin{proofEnd}
  By Lemma~\ref{lemma:inverseftilde}, 
  $\leftA$ and $\rightA$ are adjoint maps \ifff $\leftA^{-1}$ and
  $\rightA^{-1}$ are adjoints.  Assuming that
  diagram~\eqref{diag:serre} commutes, we prove commutation of
  diagram~\eqref{diag:serre2} as follows:
  \begin{center}
    \begin{tikzcd}[ampersand replacement=\&]
      B \tensor B
      \ar[bend right=15,swap]{rdd}{\leftA^{-1} \tensor B}
      \ar{rd}{\leftA^{-1}\tensor \rightA^{-1} }
      \ar[bend left=15]{rrd}{B \tensor \rightA^{-1}}
      \\
      \&A \tensor A \ar[]{d}{A \tensor \rightA}
      \ar[swap]{r}{\leftA\tensor A}
      \& B       \tensor A
      \ar{d}{\ract[A]}
      \\
     \&A \tensor B \ar{r}{\lact[A]}
      \& B
    \end{tikzcd}
  \end{center}
  A similar argument shows that commutation of
  diagram~\eqref{diag:serre2} implies that of
  diagram~\eqref{diag:serre}.
\end{proofEnd}

\vspace{-1mm}

For a \Fs $\frob$, let us denote by $\mu_B$ the diagonal of
diagram~\eqref{diag:serre2}. We next consider the magma
$(B,\mu_{B})$. Note that from the definition of $\mu_B$ and of the actions of a magma over its dual, we obtain two pairs of parallel equal arrows
\vspace{-5mm}
\begin{center}
  \begin{tikzcd}
    \node at (0,0) {\includegraphics[page=4]{serre_magma_Bserre.pdf}};
    \node at (2,-2)
    {\includegraphics[page=5]{serre_magma_Bserre.pdf}};
  \end{tikzcd}
\end{center}
\vspace{-1mm}
showing that the actions of the magma $(B,\mu_{B})$ on the dual $A$ depends on $\mu_A$ by:
\begin{align}
  \label{eq:right_action_of_B_in_term_of_muA_and_l-1}
  \lact[B] & = \mu_{A} \circ (r^{-1} \tensor A)\,,
  &
  \ract[B] & = \mu_{A} \circ (A \tensor l^{-1})\,,
\end{align}
and that $\mu_A$ is obtained from the action of $(B,\mu_B)$ over $A$
by:
\vspace{-1mm}
\begin{align}
  \label{diag:coactions}
  \begin{tikzcd}[ampersand replacement=\&]
    A \tensor A 
    \ar{r}{A \tensor \leftA}
    \ar[dotted]{dr}{\mu_{A}}
    \ar[swap]{d}{\rightA \tensor A} \& A \tensor B
    \ar{d}{\ract[B]}
    \\
    B \tensor A \ar[swap]{r}{\lact[B]} \& A\,.
  \end{tikzcd}
\end{align}

\vspace{-2mm}
We collect the properties of \Fs{s} that we want to emphasize next.
\begin{theoremEnd}
  [category=frob]
  {proposition}
  \label{lemma:frobenius_structure_various_statement}
  For a Frobenius structure $\frob$, the
  following statements hold:
  \begin{enumerate}
  \item  \label{item:firstDual}
    $(A,B,\epsilon,\mu_{A}\circ \sigma_{A,A},r,l)$ is a
      \Fs.
  \item
    \label{item:muBsemigroup}
    The magma $(B,\mu_B)$ is a semigroup.
  \item \label{item:commdiagactmu} The following diagrams commute:
    \begin{align}
      \label{diags:semipreserving}
      \begin{tikzcd}[ampersand replacement=\&]
        A \tensor A \ar{r}{\mu_{A}} \ar{d}{A \tensor l}\& A \ar{d}{l} \\
        A \tensor B  \ar{r}{\lact} \& B
      \end{tikzcd}
      &\qquad\qquad\qquad
      \begin{tikzcd}[ampersand replacement=\&]
        A \tensor A \ar{r}{\mu_{A}} \ar{d}{r \tensor A}\& A \ar{d}{r} \\
        A \tensor B  \ar{r}{\ract[A]} \& B
      \end{tikzcd}
    \end{align}
    \myVspace{-5mm}
  \item
    \label{item:inverses}
    The maps $\leftA$ and $\rightA$ are both semigroup homomorphisms
    from $(A,\mu_A)$ to $(B,\mu_B)$.
  \item  \label{item:dualFs}
    The tuple $(B,A,\epsilon \circ \sigma_{A},\mu_B,\rightA^{-1},\leftA^{-1})$ is a \Fs.
  \end{enumerate}
\end{theoremEnd}
\begin{proofEnd}
  \ref{item:firstDual} This immediately follows from the fact that the
  pairing $\epsilon \circ (A \tensor r)$ is co-associative, see
  Lemma~\ref{lemma_shift_diagram}.

  \ref{item:muBsemigroup}.  
  In view of Lemma~\ref{lemma:actions}, the following diagram commutes:
  \begin{align*}
    \begin{tikzcd}[ampersand replacement=\&,column sep=15mm]
      B \tensor  B \tensor  B \ar{rr}{B \tensor (\ract[A] \circ (B
        \tensor r^{-1}))} \ar{rd}{l^{-1} \tensor B
        \tensor r^{-1}} \ar[swap]{dd}{(\lact \circ (l^{-1}\tensor B))
        \tensor B} \& \& B
      \tensor  B \ar{d}{l^{-1}\tensor B} \\
      \& A \tensor  B \tensor  A \ar[swap]{d}{\lact \tensor A} \ar{r}{A \tensor \ract[A]} \ar{d}{} \& A \tensor  B \ar{d}{\lact}\\
      B \tensor  B \ar{r}{B \tensor l^{-1}} \& B \tensor  A \ar{r}{\ract[A]} \& B
    \end{tikzcd}
  \end{align*}
  
  \ref{item:commdiagactmu}. Using Lemma~\ref{lemma_shift_diagram}, we
  remark that the following diagram commutes:
  \begin{center}
    \begin{tikzcd}[ampersand replacement=\&]
      A\tensor A\tensor A \parrows{A \tensor \mu_{A}}{\mu_{A} \tensor
      A} \ar{d}{A \tensor A \tensor l}
    \& A\tensor A \ar{r}{A \tensor l}  \& A\tensor B \ar{d}{\epsilon} \\
      A\tensor A\tensor A \parrows{ \mu_{A} \tensor A}{A
      \tensor \lact} \& A\tensor B \ar{r}{\epsilon} \& 0
    \end{tikzcd}
  \end{center}
  We obtain therefore the equality
  $\epsilon \circ (A \tensor (\lact \circ (A \tensor l))) = \epsilon \circ
  (A \tensor (l \circ \mu_{A}))$, which, by uniqueness, implies
  $\lact \circ (A \tensor l) = l \circ \mu_{A}$.

  In a similar way we obtain that
  $\lact[A^{op}] \circ (A \tensor r) = r \circ
  \mu_{A^{op}}$. Considering that
  $\lact[A^{op}] = \ract \circ \sigma_{A,B}$, we infer
  $\ract \circ (r \tensor A) \circ \sigma_{A,A} = r \circ \mu_{A^{op}}
  = r \circ \mu_{A} \circ \sigma_{A,A}$, whence
  $\ract \circ (r \tensor A) = r \circ \mu_{A}$.

  \ref{item:inverses}. Immediate from
  diagrams~\eqref{diags:semipreserving}.
  For example:
  \begin{center}
    \begin{tikzcd}[ampersand replacement=\&]
      A \tensor A \ar{r}{\mu_{A}} \ar{d}{A \tensor l}\& A \ar{d}{l} \\
      A \tensor B \ar{d}{l \tensor B} \ar{r}{\lact} \& B \\
      A \tensor B \ar[swap]{ru}{\mu_{B}}  \\
    \end{tikzcd}
  \end{center}

  \ref{item:dualFs}.  By Lemma~\ref{lemma:DualPairSymmetry}
  $(B,A,\epsilon \circ \sigma_{B,A})$ is a dual pair. By item
  \ref{item:muBsemigroup}. of this proposition $(B,\mu_B)$ is a
  semigroup.  By Lemma~\ref{lemma:inverseftilde} $(r^{-1},l^{-1})$ are
  adjoint.  By diagram~\eqref{diag:coactions} and
  Lemma~\ref{lemma:Serre2}, the tuple
  $(B,A,\epsilon \circ \sigma_{B,A},\mu_{B},r^{-1},l^{-1})$ is a \Fs.
\end{proofEnd}

The last item of the Proposition exhibits the expected
duality. Indeed, using diagram~\eqref{diag:coactions}, it is easily
seen that the dual of
$(B,A,\epsilon \circ \sigma_{B,A},\mu_{B},r^{-1},l^{-1})$ is $\frob$
itself.  Notice that having such duality is not an obvious result,
since our definition of \Fs is not self dual. \textit{E.g.}, the dual $B$ of $A$
is not required to be a semigroup itself and the requirement that
diagram~\eqref{diag:serre} commutes is a condition on the semigroup
$(A,\mu_{A})$.

\mySkip

Finally, let us establish the connection with \abm{s}.
\begin{theoremEnd} 
  [category=frob]
{proposition} 
\label{prop:FrobStructsBrackSemigroups}
Let $\frob$ be a \Fs and define
$\pi^{\leftA}_{A} \eqdef \epsilon \circ (A \tensor \leftA)$ and
$\pi^{\rightA}_{A} \eqdef \epsilon \circ (A \tensor \rightA)$.  Then
$(A,\mu_{A},\pi^{\leftA}_{A})$ is an \abm,
$(A,\mu_{A},\pi^{\rightA}_{A})$ is a co-associative bracketed magma
and $\pi^{\leftA}_{A}$ and $\pi^{\rightA}_{A}$ are dual
pairings. Conversely, given an \abm $(A,\mu_{A},\pi_{A})$ for which
$\pi_{A}$ is a dual pairing, and given another dual paring
$(A,B,\epsilon)$, define $l$ so that
$\epsilon \circ (A \tesnor l) = \pi_{A}$, $r = \widetilde{\leftA}$,
then $\frob$ is a \Fs.
\end{theoremEnd}

\begin{proofEnd}
  By Lemma~\ref{lemma_shift_diagram}, the diagram~\eqref{diag:serre} commutes if and only if
  \begin{align*}
    \epsilon
    \circ (\mu_{A} \tensor r) & = \epsilon \circ (\mu_{A} \tensor l) \circ \sigma_{A \tensor A, A}\,.
  \end{align*}

By precomposing the equality above with the invertible
$\sigma_{A, A \tensor A}$, diagram~\eqref{diag:serre} commutes if
and only if
\begin{align*}
   \epsilon \circ (\mu_{A} \tensor l) 
   & = \epsilon
   \circ (\mu_{A} \tensor r) \circ \sigma_{A, A \tensor A} 
   = \epsilon
   \circ \sigma_{B,A} \circ (r \tensor \mu_{A})\,.
 \end{align*}
 
 Assume now that $(\leftA,\rightA)$ are adjoint.  By the previous remark,
 the diagram below commutes:
 \begin{center}
   \includegraphics[page=5]{shift_diagram}
 \end{center}
 making $\pi_A = \epsilon \circ (A \tensor \leftA)$ into an
 associative pairing. By Fact~\ref{lemma:associativeiffcoassoc},
 $\pi^{\rightA}_{A}$ is a co-associative pairing. Also, as $\epsilon$
 is a dual pairing and $\leftA$ and $\rightA$ are two isomorphisms, by
 Lemma~\ref{lemma:uniquenessDual}, $\pi^{\leftA}_{A}$ and
 $\pi^{\rightA}_{A}$ are dual pairing.
 
 Conversely, using Lemma~\ref{lemma:uniquenessDual}, $\leftA$ and
 $\rightA$ are two isomorphisms.  Notice also that if the outside
 diagram above commutes, then the rightmost commutes as well, showing
 that associativity of $\epsilon \circ (A \tensor \leftA)$ implies
 commutativity of diagram~\eqref{diag:shift}.
\end{proofEnd}

From this correspondence between \Fs{s} and \abm{s}, it is easy to find some examples from the literature.
\begin{example}
  A Frobenius algebra is a finite-dimensional $k$-algebra $(A,\mu_A)$
  coming with a non degenerate form $\pairing{-,-}$ such that
  $\pairing{xy,z} = \pairing{x,yz}$. That is, it is a bracketed
  semigroup $(A,\mu_A, \pairing{-,-})$ and $\pairing{-,-}$ is a dual
  pairing as its transpose is an isomorphism between $A$ and
  $\SA$. Hence, $(A,\SA,\ev, \mu_A,\rightA,\leftA)$ is a \Fs{} where
  $\leftA$ and $\rightA$ are defined as stated in Proposition
  \ref{prop:FrobStructsBrackSemigroups}.
\end{example}

\begin{example}
  A Frobenius algebra $(A,t,\delta,e,\mu)$ in a symmetric monoidal
  category, as defined in \cite{Hyland2004}, yields  
  an \abm $(A,\mu,t\circ \mu)$. Moreover, in \cite{Hyland2004}, it is
  noticed that $(A,A,t\circ \mu,\delta \circ e)$ is an adjunction
  which implies that $(A,A,t\circ \mu)$ is a dual pairing (see
  Example~\ref{example:adjoints}).  Whence, a \Fs{} on the semigroup
  $(A,\mu)$ is obtained  as stated in
  Proposition~\ref{prop:FrobStructsBrackSemigroups}.  
\end{example}
\begin{example}
  A similar argument shows that a Frobenius monoid
  $(A,t,\delta,e,\mu)$ in a linear distributive category as defined in
  \cite{Egger2010} yields a \Fs{}, when the category is actually
  \saut. Indeed, one obtains an \abm{} with the same construction as
  in the last example. Moreover, Theorem 3.1 of \cite{Egger2010}
  ensures that the pairing is the co-unit of a linear adjunction. That
  is $(A,A,t\circ \mu)$ is a dual pairing (see
  Example~\ref{example:adjoints}) and we obtain a \Fs{} just like
  before.
\end{example}

\begin{example}
  A Frobenius structure in the category of sets and relation is easily
  seen to amount to a tuple $(X,R,l,r)$ with $R$ a ternary relation on
  $X$ which is associative---\ie such that, for all $x,y,z,w \in X$,
  $x y R u$ and $u z R w$ (for some $u \in X$) if and only if
  $x v R w$ and $y z R v$ (for some $v \in X$)---and with $l,r$
  inverse bijections on $X$ satisfying $x y R l(z)$ if and only if
  $y z R r(x)$, for all $x,y,z \in X$.
\end{example}

%% file: adjoints.tex
\section{Nuclear objects and adjunctions}
\label{sec:adjoints}

For completeness, we state 
here that, in an autonomous category, an object is nuclear if and only
it is part of an adjunction.
This statement is actually a folklore result in the literature. For
example, it is explicitly mentioned in \cite{Rowe1988} that compact
closed categories are those autonomous categories for which every
object is nuclear, while the same categories are defined in
\cite{KellyLaplaza1980} by the property that every object is part of
an adjunction. We assume from now on that the object $\zero = I$ is the unit of the
autonomous category $\V$.

The next statement shows that, in an adjunction of the form
$(A,\SA,\eta,\epsilon)$, we can assume that the counit $\epsilon$ is
the evaluation map and, moreover, require commutativity of only one of
the two diagrams defining an adjunction. 
\begin{theoremEnd}
  [category=adjoints]
  {lemma}
  \label{lemma:adjointsAut}
  In an autonomous category the following conditions are equivalent:
  \\[-2pt]
  \twoCols[0.55]{
    \begin{enumerate}
    \item $A$ is part of an adjunction,
    \item there exists a map $\eta : I \rto \SATA$ such that
      diagram~\eqref{diag:onlyOnePart}
      commutes,
    \item there exists a map $\eta : I \rto \SATA$ such that
      $(A,\SA,\eta,\ev_{A,I})$ is an adjunction.
    \end{enumerate}
  }{
    \vspace{-10pt}
    \begin{align}
      \label{diag:onlyOnePart}
      \useDiagram[9]{sndMainthmDiags}
    \end{align}
  }
\end{theoremEnd}
\begin{proofEnd}
  If $(A,B,\upsilon,\epsilon)$ is an adjunction, then $(A,B,\epsilon)$
  is a dual pairing. By Lemma~\ref{lemma:uniquenessDual}, there exists
  an isomorphism $\psi : B \rto \SA$ such that
  $\epsilon = \ev_{A,I} \circ (A \tensor \psi)$. Define
  $\eta \eqdef ( \psi\tensor A ) \circ \upsilon$,
  we derive
  \begin{align*}
    (\ev_{A,I} \tensor A) \circ (A \tensor \eta) & =
    (\ev_{A,I} \tensor A) \circ (A \tensor \psi \tensor A) \circ (A
    \tensor \upsilon)
     \\
     &
    =  (\epsilon \tensor A)  \circ (A \tensor \upsilon) = \lambda_{A}^{-1}\circ \rho_{A} \,. 
  \end{align*}
  Suppose now that diagram~\eqref{diag:onlyOnePart} commutes. We claim
  that the diagram
  \begin{align}
    \label{diag:adjointnessOtherHalf}
    \useDiagram{sndMainthmDiags}
  \end{align}
  commutes as well, thus making $(A,\SA,\eta,\ev_{A,I})$ into an
  adjunction.
  Indeed, the transpose of $\lambda_{\SA}$ is the canonical map
  $A \tensor I \tensor \SA \rto I$ arising, up to unital isomorphisms,
  from evaluation. The commutative diagram
  \begin{center}
    \useDiagram{sndMainthmDiags}
  \end{center}
  shows that the transpose of the rightmost path in
  \eqref{diag:adjointnessOtherHalf} is the same
  canonical map.

  Finally, if $(A,\SA,\eta,\ev_{A,I})$ is an adjunction, then
  obviously $A$ is part of an an adjunction.
\end{proofEnd}

Recall 
now that an object $A$ of an autonomous category is \emph{nuclear} if
the canonical map $\mix : \SATA \rto \HAA$ is an isomorphism.
\begin{theoremEnd}
  [category=adjoints]
  {theorem}
  \label{thm:nuclear_adjoints}
  In an autonomous category, the following are equivalent:
  \begin{enumerate}
  \item $A$ is nuclear, 
  \item there exists a map $\eta : I \rto \SATA$ such that
    $\mix \circ \eta = \tr{\rho_{A}} : I \rto \HAA$,
  \item $A$ is part of an adjunction.
  \end{enumerate}
\end{theoremEnd}
\begin{proofEnd}
  If $\mix$ is invertible then define $\eta\eqdef {\mix}^{-1}\circ \tr{\rho_{A}}$. That is, 1 implies 2. 
  
  To see that 2 implies 3, observe that the relation
  $\mix \circ \eta = \tr{\rho_{A}}$ is equivalent (under
  transposition) to commutativity of
  \eqref{diag:onlyOnePart}. Therefore, by
  Lemma~\ref{lemma:adjointsAut}, $A$ is an adjoint.

  Let us argue that 3 implies 1. Suppose that $A$ is an adjoint. By
  Lemma~\ref{lemma:adjointsAut}, there exists $\eta$ such that
  $(A,\SA,\eta,\ev_{A,I})$ is an adjunction.
  Then, we define the map $\psi : \HAA \rto \SATA$ by
  \begin{align*}
    \psi & \eqdef \HAA \xrightarrow{\lambda_{\HAA}^{-1}} I \tensor \HAA 
    \xrightarrow{\eta \tensor \HAA} \SATA \tensor \HAA
    \xrightarrow{\SA \tensor \ev_{A,A}} \SATA\,.
  \end{align*}
  To see that $\mix \circ \psi$ is the identity of $\HAA$
  observe that its transpose is the evaluation map, as witnessed by the commutativity of the diagram
  \begin{center}
    \useDiagram[scale=0.8]{sndMainthmDiags}
  \end{center}
  Also, commutativity of the diagram
  \begin{align*}
    \useDiagram[scale=0.82,align=b]{sndMainthmDiags}
  \end{align*}
  shows that $\psi \circ \mix = \id_{\SATA}$. 
\end{proofEnd}

%% file: fsttheo.tex
\section{From nuclearity to \Fs{s}}
\label{sec:fsttheo}

The next two sections present our main results. First we show how
nuclearity yields \Fs{s} on the objects on the internal hom $\HAA$.

\begin{theoremEnd}
[normal]
{theorem}
  \label{thm:direct}
  If $A$ is a nuclear object in a \sautcat $\V$, then there is a map $l$ such that
  $(\HAA, \Star{\HAA},\ev_{\HAA,I},\circ,l,l)$ is a Frobenius
  structure.
\end{theoremEnd}
\begin{proofEnd}
  As mentioned in Example~\ref{example:SATAHAA_semigroup}, the object
  $\SATA$ has the semigroup structure
  $\mu_{\SATA} = \rho_{\SA} \circ (\SA \tensor \lambda_{A})\circ (\SA
  \tensor \ev_{A,I} \tensor A)$, see e.g. \cite{KrumlPaseka2008}.  It
  is easily seen that
  $\epsilon \circ (\SATA \tensor \mix) = \ev_{A,I} \circ
  \sigma_{\SA,A} \circ \mu_{\SATA}$, which immediately ensures that
  the map $\epsilon \circ (\SATA \tensor \mix)$, that is, the
  transpose of $\mix$, is associative. It is also verified that
  $\epsilon \circ (\SATA \tensor \mix) \circ \sigma_{\SATA,\SATA} =
  \epsilon \circ (\SATA \tensor \mix)$, whence $\mix$ is self-adjoint.

  It follows that, if $\mix$ is invertible, then
   $(\SATA,\HAA,\epsilon,\mu_{\SATA},\mix,\mix)$ and
  $(\HAA,\SATA,\epsilon \circ
  \sigma_{\HAA,\SATA},\mu_{\HAA},\mix^{-1},\mix^{-1})$ are
  \Fs{s}. Since $\mix$ is a semigroup homomorphism from
  $(\SATA,\mu_{\SATA})$ to $(\HAA,\circ)$, then $\mu_{\HAA}$ is the
  standard monoid structure $\circ$ induced from composition in
  $\HAA$.
  There is a canonical isomorphism $\psi : \SATA\cong \Star{\HAA}$
  such that $\epsilon \circ (\HAA \circ \psi) = \ev_{\HAA,I}$, see
  Example~\ref{example:SATAHAA}. Then, it is easily seen that
  $(\HAA,\Star{\HAA},\ev_{\HAA,I},\circ,\psi \circ \mix^{-1},\psi
  \circ \mix^{-1})$ is a \Fs.
\end{proofEnd}

Let us notice that, in view of Theorem~\ref{thm:nuclear_adjoints}
identifying nuclear objects and adjoints, the previous statement can
be seen as an instance of \cite[Corollary 3.3]{Street2004}.  This
statement admits a sort of generalization.  Since the category $\SLatt$
 has an epi-mono factorization system which lifts to the category of
quantales, we argued in \cite{deLacroixSantocanale2022} that, by
taking the appropriate quantic nucleus, one can always obtain a Girard
quantale from the image of $\mix$.  The general situation is
abstracted as follows:
\begin{theoremEnd}
  [normal]
  {theorem}
  \label{thm:mainthm}
  Let $\V$ be a \saut category with an epi-mono factorization system,
  $(A,B,\epsilon)$ be a dual pairing, and $(A,\mu_{A})$ be a semigroup
  in $\V$.  Let $f : A \rto B$ be a map, put
  $\psi_{A} \eqdef \epsilon \circ (A \tensor f)$ and suppose that
  $\psi_{A} = \psi_{A} \circ \sigma_{A,A}$.  Factor $f$ as
  $f = m \circ e$ with $e : A \rto C$ epi and $m : C \rto B$ mono. If
  $C$ is a magma with multiplication $\mu_{C}$ and $e$ is a magma
  homomorphism, then there exists a
  map 
  $g : C \rto \Star{C}$ 
  making $(C,\Star{C},\ev_{C,I},\mu_{C},g,g)$ into a Frobenius
  structure.
\end{theoremEnd}

To prove the theorem, we need a Lemma relating factorization systems
to dual pairs. If $f = m \circ e$ with $e$ epi and $m$ mono, then we
denote by $Im(f)$ the codomain of $e$.
\begin{theoremEnd}
  [normal]
  {lemma}
  \label{lemma:factorization}
  Let $\V$ be \saut with an epi-mono factorization system. If $(A,B)$
  is a dual pair and $f : A \rto B$ then $(Im(f),Im(\Tilde{f}))$ is a
  dual pair. 
\end{theoremEnd}
\begin{proofEnd}
  As from Proposition~\ref{prop:charDualPair}.\ref{item:dualABstars},
  let $\phi : A \rto \SB$ and $\psi : B \rto \SA$ be isomorphisms such
  that $\phi = \Perp{\psi}$.  We pretend that, up to these
  isomorphisms, $\Tilde{f} $ equals $\Star{f}$, as stated below.

\vspace{-1mm}

  \begin{theoremEnd}[normal]{claim}
    We have $\psi \circ \Tilde{f} = \Star{f} \circ \phi$.
  \end{theoremEnd}
  \begin{theoremEnd}
    [category=fsttheo, all end]
    {claim} \label{lemmaftildefperp}
    The following
    diagram commutes:
    \begin{center}
      \begin{tikzcd}[ampersand replacement=\&]
        A \ar{r}{\Tilde{f}} 
        \ar{d}{\phi} 
        \& B
        \ar{d}{\psi} 
        \\
        \SB 
        \ar{r}{\Star{f}} 
        \& \SA
      \end{tikzcd}
    \end{center}  
  \end{theoremEnd}
  \begin{proofEnd}
  By transposing $\psi \circ \Tilde{f}$, we obtain the upper leg in
  the following diagram:
  \begin{center}
    \begin{tikzcd}[column sep=15mm,ampersand replacement=\&]
      A \tensor A
      \ar{d}{\sigma_{A,A}} 
       \ar{r}{A \tensor \Tilde{f}} \& 
       A \tensor B \ar{r}{A     \tensor \psi} 
       \ar{rd}{\epsilon} 
       \& A \tensor \SA \ar{d}{\ev_{A,0}} 
       \\
      A \tensor A 
      \ar{r}{A \tensor f}
      \& A \tensor B 
      \ar{d}{A \tensor    \psi}
       \ar{r}{\epsilon} 
       \& 0 \\
      \& A \tensor \SA 
      \ar[swap]{ru}{\ev_{A,0}} 
    \end{tikzcd}
  \end{center}
  The lower leg of this diagram being the transpose of
  $\Perp{(\psi \circ f)}$, we deduce
  $\psi \circ \Tilde{f} = \Perp{(\psi \circ f)}$.  By the naturality
  of the operation $\Perp{\fun}$,
  we have
  \begin{align*}
    \Perp{(\Star{f} \circ \phi)} & = \Perp{\phi} \circ f  = \psi \circ f
  \end{align*}
  whence $\Star{f} \circ \phi  =  \Perp{(\psi \circ f)} = 
    \psi \circ \Tilde{f}$.
  \end{proofEnd}
  Let now $(e,m)$ and $(\Tilde{e},\Tilde{m})$ be epi-mono factorizations
  of $f$ and $\Tilde{f}$, respectively.
  By applying the duality functor $\Star{\fun}$,
  $(\Star{m}, \Star{e})$ (resp.
  $(\Star{\Tilde{m}}, \Star{\Tilde{e}})$) is an epi-mono factorization
  of $\Star{f}$ (resp. $\Star{\Tilde{f}}$). 
  Therefore, we obtain two different epi-mono factorizations of
  $\psi \circ \Tilde{f} = \Star{f}\circ \phi$ (resp.
  $\psi \circ f = \Star{\Tilde{f}}\circ \phi$), as follows:
  \begin{center}
    \begin{tikzcd}[ampersand replacement=\&, row sep = 4mm]
      A 
      \ar{rd}{\Tilde{e}}\ar{rr}{\Tilde{f}}
       \ar{ddd}{\phi}
       \&\& B 
       \ar{ddd}{\psi}\\
      \& Im(\Tilde{f})
       \ar{ru}{\Tilde{m}} 
       \ar[dotted]{d}{\chi}
        \\
      \& \Star{Im(f)} 
       \ar{rd}{\Star{e}} \& 
       \\
      \SB \ar{ru}{\Star{m}}
       \ar{rr}{\Star{f}}
        \&\& \SA
    \end{tikzcd}
    \qquad\qquad
    \begin{tikzcd}[ampersand replacement=\&, row sep = 4mm]
      A 
      \ar{rd}{e}
      \ar{rr}{f} 
      \ar{ddd}{\phi}
      \&\& B 
      \ar{ddd}{\psi}
      \\
      \& Im(f)
       \ar{ru}{m}
       \ar[dotted]{d}{\xi} 
       \\
      \& \Star{Im(\Tilde{f})}
        \ar{rd}{\Star{\Tilde{e}}} 
         \\
      \SB 
      \ar{ru}{\Star{\Tilde{m}}}
       \ar{rr}{\Star{\Tilde{f}}} 
       \&\& \SA
    \end{tikzcd}
  \end{center}
  and therefore isomorphisms $\chi : Im(\Tilde{f}) \rto \Star{Im(f)}$
  and $\xi : Im(f) \rto \Star{Im(\Tilde{f})}$. In order to conclude
  that we have a dual pair, we need to argue that $\xi= \Perp{\chi}$.
  To this goal, compute as follows:
  \begin{align*}
    \Perp{(\chi \circ \Tilde{e})} \circ e
    & = \Perp{(\Star{e} \circ \chi \circ \Tilde{e})} 
     = \Perp{(\Star{f} \circ \phi)} 
     =\Perp{\phi} \circ f 
    = \psi\circ f 
    = \Star{\Tilde{e}} \circ \xi \circ e\,.
  \end{align*}
  From this, considering that $e$ is epi, it follows that
  $\Perp{(\chi \circ \Tilde{e})} = \Star{\Tilde{e}} \circ \xi$.  By
  applying the operator $\Perp{\fun}$ to both sides of this equality,
  we obtain $\chi \circ \Tilde{e}  = \Perp{(\Star{\Tilde{e}} \circ \xi)} =
  \Perp{\xi} \circ \Tilde{e}$,
  from which the desired equality $\chi =  \Perp{\xi}$ follows.
\end{proofEnd}

\begin{proof}
  [Proof of Theorem~\ref{thm:mainthm}] Let $A,B$ be dual and let
  $f : A \rto B$ be such that $\Tilde{f} = f$. Let $e : A \rto C$ and
  $m : C \rto B$ be a factorization of
  $f$. Lemma~\ref{lemma:factorization} exhibits $(C,C)$ as a dual pair
  with isomorphism $\chi,\xi : C \rto \SC$ such that
  $\Perp{\chi} = \xi = \chi$. Define therefore the pairing
  $\pi_{C} \eqdef \ev_{C,I} \circ (C \tensor \chi) : C \tensor C \rto I$
  and observe that it is symmetric.  The diagram in the proof of
  Lemma~\ref{lemma:factorization} also exhibits the equality
  $\psi \circ f = \Star{e} \circ \chi \circ e : A \rto \SA$.
  Transposing this relation, we obtain the pairing 
  \begin{align*}
    \pi_{A} & = \trdown{(\psi \circ f)} = \trdown{(\Star{e} \circ \chi
      \circ e)} = \trdown{\chi} \circ (e\tensor e) = \pi_{C} \circ (e\tensor e)\,.
  \end{align*}
  Thus, assuming that $f : A \rto B$ is a semigroup homomorphism
  factoring as $f = m \circ e$ with $e : (A,\mu_{A}) \rto (C,\mu_{C})$
  a magma homomorphisms, we have that
  $e : (A,\mu_{A},\pi_{A}) \rto (C,\mu_{C},\pi_{C})$ is a bracketed
  magma homomorphism. As argued in
  Proposition~\ref{prop:quotientSemigroup}, $(C,\mu_{C},\pi_{C})$ is
  an associative bracketed magma,
  and, as mentioned in
  Proposition~\ref{prop:FrobStructsBrackSemigroups}, this is enough
  to have a \Fs on the dual pair $(C,\SC)$.
\end{proof}

%% file: sndtheo.tex
\section{From \Fs{s} to nuclear objects}
\label{sec:sndtheo}

We finally give the converse of Theorem~\ref{thm:direct}. However, to do so we
impose additional conditions to the ambient \saut category $\V$.

\begin{definition}
  We say that an object $A$ of a monoidal category $\V$ is
  \emph{\pseudoaffine} if the tensor unit $I$ embeds into $A$ as a
  retract. If every object of $\V$ which is not terminal nor initial
  is \pseudoaffine, then $\V$ is said to be \emph{\pseudoaffine}.
\end{definition}
For instance, the category \SLatt is \pseudoaffine. This property was
actually used in \cite{S_ACT2020} to prove that if $\HLL$ is endowed
with a Frobenius quantale structure, then $L$ is \cd. 
Before going there, let us introduce some useful observations.

\textEnd[category=sndtheo]{
  If $A$ is \pseudoaffine, then we can pick $p : I \rto A$ and $c : A \rto I$
such that $c \circ p = \id_{I}$. We shall assume that any \pseudoaffine $A$
comes with a canonical choice of such $p$ and $c$.  Then, for $A$
\pseudoaffine, for each object $X$ we can define
\begin{align*}
  p^{\rho}_{X} & : X \rto X
  \tensor A \,, & p^{\rho}_{X} & \eqdef (X \tensor p)\circ \rho^{-1}_{X}\,,\\
  c^{\rho}_{X} & : X
  \tensor A \rto X & c^{\rho}_{X} & \eqdef \rho_{X} \circ (X \tensor c)\,,\\
  p^{\ell}_{X} & : X \rto A
  \tensor X & p^{\ell}_{X} & \eqdef (p \tensor X) \circ \lambda^{-1}_{X}\,,\\
  c^{\ell}_{X} & : A \tensor X \rto X & c^{\ell}_{X} & \eqdef \lambda_{X}
  \circ (c \tensor X)\,.
\end{align*}
Clearly these collections of arrows are natural in $X$ and the
following relations hold:
\begin{align*}
  c^{\rho} \circ p^{\rho} & = \id\,, & c^{\ell} \circ p^{\ell} & =
  \id\,.
\end{align*}
Notice $p^{\rho}_{X \tensor Y} = X \tensor p^{\rho}_{Y}$ and similar
relations holds for $p^{\ell},c^{\rho},c^{\ell}$. Therefore, if $A$
and $C$ are \pseudoaffine, then the following diagram commutes (using
naturality):
\begin{center}
  \useDiagram[1]{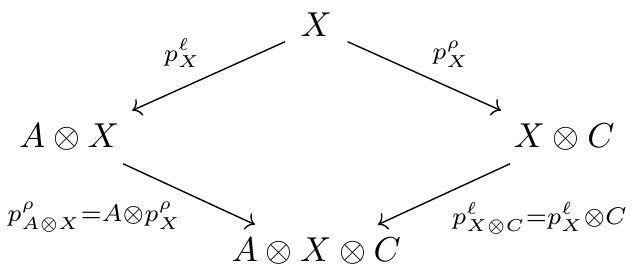}
\end{center}
Similar relations hold for $c^{\ell}$ and $c^{\rho}$.
}

\begin{theoremEnd}
  [category=sndtheo]
  {lemma}
  Tensoring with a \pseudoaffine object yields a faithful
  functor.
\end{theoremEnd}
\begin{proofEnd}
  Let $A$ be \pseudoaffine and suppose that
  $f,g : X \rto Y$ and
  consider the following diagram:
  \begin{center}
    \useDiagram[1]{affineDiags}
  \end{center}
  If $f \tensor A = g \tensor A : X \tensor A \rto Y \tensor A$, then
  also $p^{\rho}_{Y} \circ f = p^{\rho}_{Y} \circ g$ and then $f = g$,
  since $p^{\rho}_{Y}$ is monic.
\end{proofEnd}

\textEnd[category=sndtheo]{
Notice that the previous proof shows that faithful functors are closed
under super-functors. 
 }

\begin{theoremEnd}
  [category=sndtheo]
  {proposition}
  \label{prop:orthogonal}
  Identity morphisms of \pseudoaffine objects are orthogonal to each other,
  in the following sense: if $A,C$ are \pseudoaffine,
  $f : A \tensor B_{0} \rto A \tensor B_{1}$,
  $g : B_{0} \tensor C \rto B_{1} \tensor C$, and
  $f \tensor C = A \tensor g : A \tensor B_{0} \tensor C \rto A
  \tensor B_{1} \tensor C$, then there exists $h : B_{0} \rto B_{1}$
  such that $f= A \tensor h$ and $g = h \tensor C$.
\end{theoremEnd}
\begin{proofEnd}
  First, using that $C$ is \pseudoaffine, we have
  \begin{align*}
    f & = c^{\rho}_{A \tensor B_{0}} \circ (f \tensor C) \circ
    p^{\rho}_{A \tensor B_{0}} = c^{\rho}_{A \tensor B_{1}} \circ (A
    \tensor g) \circ
    p^{\rho}_{A \tensor B_{0}} \\
    & \quad = (A \tensor c^{\rho}_{B_{1}}) \circ (A \tensor g) \circ A
    \tensor (p^{\rho}_{B_{0}}) = A \tensor (c^{\rho}_{B_{0}} \circ g
    \circ p^{\rho}_{B_{0}}) 
  \end{align*}

Using this equation and various properties of naturality, we obtain
  \begin{align*}
    c^{\ell}_{B_{1}} \circ f \circ p^{\ell}_{B_{0}} & = c^{\ell}_{B_{1}} \circ ( A \tensor (c^{\rho}_{B_{0}} \circ g
    \circ p^{\rho}_{B_{0}}) ) \circ p^{\ell}_{B_{0}}\\
    & =
    c^{\ell}_{B_{1}} \circ
    (A \tensor c^{\rho}_{B_{1}}) \circ (A \tensor g) \circ (A \tensor p^{\rho}_{B_{0}}) \circ
    p^{\ell}_{B_{0}}\\
    & =
    c^{\ell}_{B_{1}} \circ
    c^{\rho}_{A \tensor B_{1}}\circ (A \tensor g) \circ p^{\rho}_{A \tensor B_{0}} \circ
    p^{\ell}_{B_{0}}\\
    & =
    c^{\rho}_{B_{1}} \circ
    c^{\ell}_{B_{1}\tensor C}  \circ (A \tensor g) \circ p^{\ell}_{B_{0}\tensor C} \circ
    p^{\rho}_{B_{0}}  \\
    & = c^{\rho}_{B_{1}} \circ  g \circ  p^{\rho}_{B_{0}}\,.
  \end{align*}

  Therefore, we can define
  $h \eqdef c^{\ell}_{B_{1}} \circ f \circ p^{\ell}_{B_{0}} =
  c^{\rho}_{B_{1}} \circ g \circ p^{\rho}_{B_{0}}$
  which complete the proof of the statement.
\end{proofEnd}

The following is the main result in this section.
\begin{theoremEnd}
  [normal]
  {theorem}
  \label{thm:FBadjoint} If $A$ is \pseudoaffine and the semigroup
  $(\HAA,\circ)$ can be completed to a \Fs, then $A$ is part of an
  adjunction.
\end{theoremEnd}
\begin{proofEnd}We can suppose that the dual of $\HAA$ is $\SATA$.  The dual
  multiplication $\mu_{\SATA }: \SATA \tensor \SATA \rto \SATA$
  arises
  from a map $f : \SATA \rto \HAA$ \hfill as the diagonal of
  \\[-6pt]
  \twoCols[0.44]{
    the diagram on the right.
    Since $A$ is \pseudoaffine, then, by
    Proposition~\ref{prop:orthogonal}, we necessarily have
    \begin{align*}
      \mu_{\SATA}  = (\SA \tensor \lambda_{A}) \circ (\SA \tensor \epsilon
      \tensor A) 
    \end{align*}  
    for a map $\epsilon : \ATSA \rto I$.
  }{
    \myVspace{-20pt}
    \begin{center}
      \includegraphics[page=6]{sndMainthmDiags}
    \end{center}}\\[2pt]
  Moreover, since $(\HAA,\circ)$
  is unital and $f$ is an isomorphism, $\mu_{\SATA}$ has a unit
  $\eta : I \rto \SATA$, which means that the following diagrams
  commute:
  \begin{center}
    \myVspace{-2mm}
    \includegraphics[page=7]{sndMainthmDiags}  
  \end{center}
  \myVspace{-2mm}
  Since tensoring with $A$ and $\SA$ is faithful, commutativity of the
  above diagrams is equivalent to the equalities
  \myVspace{-2mm}
  \begin{align*}
    \lambda_{A}\circ (\epsilon \tensor A) \circ (A \tensor \eta)
    \circ \rho^{-1}_{A}
    & = id_{A}\,, & id_{\SA}& =  \rho_{A}\circ (\SA \tensor \epsilon) \circ (\eta \tensor \SA)
    \circ \lambda^{-1}_{A}\,,
  \end{align*}
  \myVspace{-2mm}
  showing that $(A,\SA,\eta,\epsilon)$ is an adjunction.
\end{proofEnd}

In view of Theorem~\ref{thm:nuclear_adjoints}, we can restate
Theorem~\ref{thm:FBadjoint} as follows:
\begin{theoremEnd}
  [normal] {theorem}
  \label{thm:converse}
  If $A$ is \pseudoaffine and $(\HAA, \circ)$ carries a Frobenius
  structure, then $A$ is nuclear.
\end{theoremEnd}

%% file: conclusions.tex
\section{Discussion and future work}
\label{sec:conclusions}

The work~\cite{DePaivaSchalk2004} provides a wide playground where to
test the strength of the results presented in this paper. For example,
for a commutative Girard quantale $Q$, the category $\QSets$, whose
objects are the pairs $(\alpha,X)$ with $\alpha : X \rto Q$ and whose
morphisms $(\alpha,X) \rto (\beta,Y)$ are the relations
$R \subseteq X \times Y$ such that $xR y$ implies
$\alpha(x) \leq \beta(y)$, is \saut.  If the relation $1 = \Rneg{1}$
holds in $Q$, then the unit of the monoidal structure is a dualizing
object. It is not difficult to characterize \paffine and nuclear
objects in this category.
It turns out that if a monoid on $\homm{(\alpha,X)}{(\alpha,X)}$ can
be endowed with a Frobenius structure, then $(\alpha,X)$ is nuclear,
even when $(\alpha,X)$ is not \paffine.  That is, this sort of model
shows that, in Theorem~\ref{thm:FBadjoint}, the assumption that an
object is \paffine is not necessary. We are leaving it open whether
this assumption can be eliminated.

The notion of nuclearity was originally conceived for Banach spaces
\cite{Grothendieck1955}. In \cite{deLacroixSantocanale2022} we have
shown how to construct unitless Girard quantales from nuclear endomaps
(trace class operators) of an infinite dimensional Hilbert space.  It
is natural to ask how the results presented in this paper may be
extended to categories of Banach spaces and other similar categories
(\eg the category of Hilbert spaces). Most of these categories are
symmetric monoidal; however, they are not \saut{} and, in the case of
Hilbert spaces and continuous linear maps, not even autonomous.  Even
though the definitions of dual pair and of \Fs{} only require the
environment category to be symmetric monoidal, a closer look at our
results exhibits their dependency on $\ast$-autonomy.
For example, we have used in Theorem~\ref{thm:mainthm} the fact that,
for $m$ mono, $\Star{m}$ is epi. In categories of Banach spaces, due
to the Hanh-Banach theorem, the theorem might still hold if $m$ is a
\emph{regular} mono, that is, an isometry. However, when considering
the algebra of trace class operators---which, given the analogy with
the tight endomaps of \cite{deLacroixSantocanale2022}, is the natural
candidate where to apply Theorem~\ref{thm:mainthm}---the relevant map
$m$ is not an isometry.  We also have used in
Theorem~\ref{thm:converse} and elsewhere that $(\SATA,\HAA)$ is a dual
pair.  This fails in autonomous categories, for example, if $A$ is the
reflexive Banach space $\ell_{p}$ with $1 < p < \infty$, then the
tensor $\SATA$ is no longer reflexive, see \cite{AriasFarmer1996}.
One might restrict to categories of finite dimensional Banach space,
which are \saut. However, these categories turn out to be
uninteresting: if bounded linear maps are taken as morphisms, then all
object are nuclear, and if we restrict maps to linear contractions
(those linear maps for which $\norm{f} \leq 1$), then the only nuclear
objects are isomorphic to the monoidal unit \cite[\S 4]{Rowe1988}.

Future research will continue investigating applications of these
theorems in concrete \saut categories. A main difficulty here will be
finding adequate characterizations of nuclear and \paffine objects.
We shall also investigate how to relax $\ast$-autonomy, which might
yield results on categories appealing for a wider community of computer
scientists and mathematicians---and a step, on our side, towards
categorical models of quantum computing.

The considerations just developed suggest, on the other hand, the
existence of a close interdependence between provability models of
classical linear logic (Frobenius quantales and, in a wider sense,
Frobenius structures) and models of proofs of this logic (the \saut
categories). It is our desire to investigate further whether there is
any relation between these structures, similarly to what happens for
intuitionistic logic with the Heyting algebra of truth values of a
topos. Steps in this direction have already been taken, see \eg
\cite{DePaivaSchalk2004,Jacobs2015}.

%% file: 0.bbl
\begin{thebibliography}{10}
\providecommand{\bibitemdeclare}[2]{}
\providecommand{\surnamestart}{}
\providecommand{\surnameend}{}
\providecommand{\urlprefix}{Available at }
\providecommand{\url}[1]{\texttt{#1}}
\providecommand{\href}[2]{\texttt{#2}}
\providecommand{\urlalt}[2]{\href{#1}{#2}}
\providecommand{\doi}[1]{doi:\urlalt{http://dx.doi.org/#1}{#1}}
\providecommand{\eprint}[1]{arXiv:\urlalt{https://arxiv.org/abs/#1}{#1}}
\providecommand{\bibinfo}[2]{#2}

\bibitemdeclare{article}{ABP1999}
\bibitem{ABP1999}
\bibinfo{author}{Samson \surnamestart Abramsky\surnameend},
  \bibinfo{author}{Richard \surnamestart Blute\surnameend} \&
  \bibinfo{author}{Prakash \surnamestart Panangaden\surnameend}
  (\bibinfo{year}{1999}): \emph{\bibinfo{title}{Nuclear and trace ideals in
  tensored {{\(^*\)}}-categories}}.
\newblock {\sl \bibinfo{journal}{J. Pure Appl. Algebra}}
  \bibinfo{volume}{143}(\bibinfo{number}{1-3}), pp. \bibinfo{pages}{3--47},
  \doi{10.1016/S0022-4049(98)00106-6}.

\bibitemdeclare{incollection}{AbramskyHeunen2012}
\bibitem{AbramskyHeunen2012}
\bibinfo{author}{Samson \surnamestart Abramsky\surnameend} \&
  \bibinfo{author}{Chris \surnamestart Heunen\surnameend}
  (\bibinfo{year}{2012}): \emph{\bibinfo{title}{H*-algebras and nonunital
  {F}robenius algebras: first steps in infinite-dimensional categorical quantum
  mechanics}}.
\newblock In: {\sl \bibinfo{booktitle}{Mathematical foundations of information
  flow}}, {\sl \bibinfo{series}{Proc. Sympos. Appl.
  Math.}}~\bibinfo{volume}{71}, \bibinfo{publisher}{Amer. Math. Soc.,
  Providence, RI}, pp. \bibinfo{pages}{1--24}, \doi{10.1090/psapm/071/599}.
\newblock \urlprefix\url{https://doi.org/10.1090/psapm/071/599}.

\bibitemdeclare{article}{AbramskyVickers1993}
\bibitem{AbramskyVickers1993}
\bibinfo{author}{Samson \surnamestart Abramsky\surnameend} \&
  \bibinfo{author}{Steven \surnamestart Vickers\surnameend}
  (\bibinfo{year}{1993}): \emph{\bibinfo{title}{Quantales, Observational Logic
  and Process Semantics}}.
\newblock {\sl \bibinfo{journal}{Math. Struct. Comput. Sci.}}
  \bibinfo{volume}{3}(\bibinfo{number}{2}), pp. \bibinfo{pages}{161--227},
  \doi{10.1017/S0960129500000189}.
\newblock \urlprefix\url{https://doi.org/10.1017/S0960129500000189}.

\bibitemdeclare{article}{ACS1998}
\bibitem{ACS1998}
\bibinfo{author}{Haroun \surnamestart Amira\surnameend}, \bibinfo{author}{Bob
  \surnamestart Coecke\surnameend} \& \bibinfo{author}{Isar \surnamestart
  Stubbe\surnameend} (\bibinfo{year}{1998}): \emph{\bibinfo{title}{How
  Quantales Emerge by Introducing Induction within the Operational Approach.}}
\newblock {\sl \bibinfo{journal}{Phys. Acta}} \bibinfo{volume}{71}, pp.
  \bibinfo{pages}{554--572}.

\bibitemdeclare{article}{AriasFarmer1996}
\bibitem{AriasFarmer1996}
\bibinfo{author}{Alvaro \surnamestart Arias\surnameend} \&
  \bibinfo{author}{Jeff~D. \surnamestart Farmer\surnameend}
  (\bibinfo{year}{1996}): \emph{\bibinfo{title}{On the structure of tensor
  products of {$l_p$}-spaces}}.
\newblock {\sl \bibinfo{journal}{Pacific J. Math.}}
  \bibinfo{volume}{175}(\bibinfo{number}{1}), pp. \bibinfo{pages}{13--37}.
\newblock \urlprefix\url{http://projecteuclid.org/euclid.pjm/1102364179}.

\bibitemdeclare{book}{Barr1979}
\bibitem{Barr1979}
\bibinfo{author}{Michael \surnamestart Barr\surnameend} (\bibinfo{year}{1979}):
  \emph{\bibinfo{title}{{$\ast$}-autonomous categories}}.
\newblock {\sl \bibinfo{series}{Lecture Notes in Mathematics}}
  \bibinfo{volume}{752}, \bibinfo{publisher}{Springer, Berlin},
  \doi{10.1007/BFb0064582}.

\bibitemdeclare{article}{BCS2000}
\bibitem{BCS2000}
\bibinfo{author}{R.~F. \surnamestart Blute\surnameend},
  \bibinfo{author}{J.~R.~B. \surnamestart Cockett\surnameend} \&
  \bibinfo{author}{R.~A.~G. \surnamestart Seely\surnameend}
  (\bibinfo{year}{2000}): \emph{\bibinfo{title}{Feedback for linearly
  distributive categories: {Traces} and fixpoints}}.
\newblock {\sl \bibinfo{journal}{J. Pure Appl. Algebra}}
  \bibinfo{volume}{154}(\bibinfo{number}{1-3}), pp. \bibinfo{pages}{27--69},
  \doi{10.1016/S0022-4049(99)00180-2}.

\bibitemdeclare{article}{Jacobs2015}
\bibitem{Jacobs2015}
\bibinfo{author}{Kenta \surnamestart Cho\surnameend}, \bibinfo{author}{Bart
  \surnamestart Jacobs\surnameend}, \bibinfo{author}{Bas \surnamestart
  Westerbaan\surnameend} \& \bibinfo{author}{Abraham \surnamestart
  Westerbaan\surnameend} (\bibinfo{year}{2015}): \emph{\bibinfo{title}{An
  Introduction to Effectus Theory}}.
\newblock {\sl \bibinfo{journal}{CoRR}} \bibinfo{volume}{abs/1512.05813}.
\newblock \eprint{1512.05813}.

\bibitemdeclare{article}{CHS2006}
\bibitem{CHS2006}
\bibinfo{author}{J.~R.~B. \surnamestart Cockett\surnameend},
  \bibinfo{author}{M.~\surnamestart Hasegawa\surnameend} \&
  \bibinfo{author}{R.~A.~G. \surnamestart Seely\surnameend}
  (\bibinfo{year}{2006}): \emph{\bibinfo{title}{Coherence of the double
  involution on {$\ast$}-autonomous categories}}.
\newblock {\sl \bibinfo{journal}{Theory Appl. Categ.}} \bibinfo{volume}{17},
  pp. \bibinfo{pages}{No. 2, 17--29}.

\bibitemdeclare{article}{Cockett99introductionto}
\bibitem{Cockett99introductionto}
\bibinfo{author}{J.~R.~B. \surnamestart Cockett\surnameend},
  \bibinfo{author}{J.~\surnamestart Koslowski\surnameend} \&
  \bibinfo{author}{R.~A.~G. \surnamestart Seely\surnameend}
  (\bibinfo{year}{2000}): \emph{\bibinfo{title}{Introduction to linear
  bicategories}}.
\newblock {\sl \bibinfo{journal}{Math. Struct. Comput. Sci.}}
  \bibinfo{volume}{10}(\bibinfo{number}{2}), pp. \bibinfo{pages}{165--203},
  \doi{10.1017/S0960129520003047}.

\bibitemdeclare{article}{CockettSeely1997}
\bibitem{CockettSeely1997}
\bibinfo{author}{J.R.B. \surnamestart Cockett\surnameend} \&
  \bibinfo{author}{R.A.G. \surnamestart Seely\surnameend}
  (\bibinfo{year}{1997}): \emph{\bibinfo{title}{Weakly distributive
  categories}}.
\newblock {\sl \bibinfo{journal}{Journal of Pure and Applied Algebra}}
  \bibinfo{volume}{114}(\bibinfo{number}{2}), pp. \bibinfo{pages}{133--173}.

\bibitemdeclare{article}{CCP2021}
\bibitem{CCP2021}
\bibinfo{author}{Robin \surnamestart Cockett\surnameend}, \bibinfo{author}{Cole
  \surnamestart Comfort\surnameend} \& \bibinfo{author}{Priyaa~V. \surnamestart
  Srinivasan\surnameend} (\bibinfo{year}{2021}): \emph{\bibinfo{title}{Dagger
  linear logic for categorical quantum mechanics}}.
\newblock {\sl \bibinfo{journal}{Log. Methods Comput. Sci.}}
  \bibinfo{volume}{17}(\bibinfo{number}{4}), p.~\bibinfo{pages}{73}.
\newblock \urlprefix\url{lmcs.episciences.org/8716}.
\newblock \bibinfo{note}{Id/No 8}.

\bibitemdeclare{article}{EggerKruml2010}
\bibitem{EggerKruml2010}
\bibinfo{author}{J.~M. \surnamestart Egger\surnameend} \&
  \bibinfo{author}{David \surnamestart Kruml\surnameend}
  (\bibinfo{year}{2010}): \emph{\bibinfo{title}{Girard {Couples} of
  {Quantales}}}.
\newblock {\sl \bibinfo{journal}{Applied Categorical Structures}}
  \bibinfo{volume}{18}(\bibinfo{number}{2}), pp. \bibinfo{pages}{123--133},
  \doi{10.1007/s10485-008-9138-3}.
\newblock \urlprefix\url{https://doi.org/10.1007/s10485-008-9138-3}.

\bibitemdeclare{article}{Egger2010}
\bibitem{Egger2010}
\bibinfo{author}{J.M. \surnamestart Egger\surnameend} (\bibinfo{year}{2010}):
  \emph{\bibinfo{title}{The {F}robenius relations meet linear distributivity.}}
\newblock {\sl \bibinfo{journal}{Theory and Applications of Categories
  [electronic only]}} \bibinfo{volume}{24}, pp. \bibinfo{pages}{25--38}.
\newblock \urlprefix\url{http://eudml.org/doc/223263}.

\bibitemdeclare{book}{EGHK2018}
\bibitem{EGHK2018}
\bibinfo{author}{Patrik \surnamestart Eklund\surnameend},
  \bibinfo{author}{Javier \surnamestart Guti\'{e}rrez~Garc{i}a\surnameend},
  \bibinfo{author}{Ulrich \surnamestart H\"{o}hle\surnameend} \&
  \bibinfo{author}{Jari \surnamestart Kortelainen\surnameend}
  (\bibinfo{year}{2018}): \emph{\bibinfo{title}{Semigroups in complete
  lattices}}.
\newblock {\sl \bibinfo{series}{Developments in
  Mathematics}}~\bibinfo{volume}{54}, \bibinfo{publisher}{Springer, Cham},
  \doi{10.1007/978-3-319-78948-4}.

\bibitemdeclare{book}{GJKO}
\bibitem{GJKO}
\bibinfo{author}{Nikolaos \surnamestart Galatos\surnameend},
  \bibinfo{author}{Peter \surnamestart Jipsen\surnameend},
  \bibinfo{author}{Tomasz \surnamestart Kowalski\surnameend} \&
  \bibinfo{author}{Hiroakira \surnamestart Ono\surnameend}
  (\bibinfo{year}{2007}): \emph{\bibinfo{title}{Residuated Lattices: An
  Algebraic Glimpse at Substructural Logics}}.
\newblock {\sl \bibinfo{series}{Studies in Logic and the Foundations of
  Mathematics}} \bibinfo{volume}{151}, \bibinfo{publisher}{Elsevier},
  \doi{10.1016/S0049-237X(07)80005-X}.

\bibitemdeclare{incollection}{GG2019}
\bibitem{GG2019}
\bibinfo{author}{Stefano \surnamestart Gogioso\surnameend} \&
  \bibinfo{author}{Fabrizio \surnamestart Genovese\surnameend}
  (\bibinfo{year}{2019}): \emph{\bibinfo{title}{Quantum field theory in
  categorical quantum mechanics}}.
\newblock In: {\sl \bibinfo{booktitle}{Proceedings of the 15th international
  conference on quantum physics and logic, QPL'18, Halifax, Canada, June 3--7,
  2018}}, \bibinfo{publisher}{Waterloo: Open Publishing Association (OPA)}, pp.
  \bibinfo{pages}{163--177}.
\newblock \urlprefix\url{eptcs.web.cse.unsw.edu.au/paper.cgi?QPL2018.9}.

\bibitemdeclare{inproceedings}{SAN-2018-RAMICS}
\bibitem{SAN-2018-RAMICS}
\bibinfo{author}{Maria~Jo{\~a}o \surnamestart Gouveia\surnameend} \&
  \bibinfo{author}{Luigi \surnamestart Santocanale\surnameend}
  (\bibinfo{year}{2018}): \emph{\bibinfo{title}{Mix $\star$-autonomous
  quantales and the continuous weak order}}.
\newblock In \bibinfo{editor}{Jules \surnamestart Desharnais\surnameend},
  \bibinfo{editor}{Walter \surnamestart Guttmann\surnameend} \&
  \bibinfo{editor}{Stef \surnamestart Joosten\surnameend}, editors: {\sl
  \bibinfo{booktitle}{RAMiCS 2018}}, {\sl \bibinfo{series}{Lecture Notes in
  Computer Science}} \bibinfo{volume}{11194}, \bibinfo{publisher}{Springer,
  Cham}, pp. \bibinfo{pages}{184--201}.
\newblock \urlprefix\url{https://doi.org/10.1007/978-3-030-02149-8_12}.

\bibitemdeclare{article}{SAN-2021-JPAA}
\bibitem{SAN-2021-JPAA}
\bibinfo{author}{Maria~Jo{\~a}o \surnamestart Gouveia\surnameend} \&
  \bibinfo{author}{Luigi \surnamestart Santocanale\surnameend}
  (\bibinfo{year}{2021}): \emph{\bibinfo{title}{The continuous weak order}}.
\newblock {\sl \bibinfo{journal}{J. Pure Appl. Algebra}}
  \bibinfo{volume}{225}(\bibinfo{number}{2}), p.~\bibinfo{pages}{37},
  \doi{10.1016/j.jpaa.2020.106472}.
\newblock \bibinfo{note}{Id/No 106472}.

\bibitemdeclare{article}{Grothendieck1955}
\bibitem{Grothendieck1955}
\bibinfo{author}{Alexandre \surnamestart Grothendieck\surnameend}
  (\bibinfo{year}{1955}): \emph{\bibinfo{title}{Produits tensoriels
  topologiques et espaces nucl\'{e}aires}}.
\newblock {\sl \bibinfo{journal}{Mem. Amer. Math. Soc.}} \bibinfo{volume}{16},
  p. \bibinfo{pages}{Chapter 1: 196 pp.; Chapter 2: 140}.

\bibitemdeclare{article}{HiggsRowe1989}
\bibitem{HiggsRowe1989}
\bibinfo{author}{D.~A. \surnamestart Higgs\surnameend} \&
  \bibinfo{author}{K.~A. \surnamestart Rowe\surnameend} (\bibinfo{year}{1989}):
  \emph{\bibinfo{title}{Nuclearity in the category of complete semilattices}}.
\newblock {\sl \bibinfo{journal}{J. Pure Appl. Algebra}}
  \bibinfo{volume}{57}(\bibinfo{number}{1}), pp. \bibinfo{pages}{67--78},
  \doi{10.1016/0022-4049(89)90028-5}.

\bibitemdeclare{inproceedings}{Hyland2004}
\bibitem{Hyland2004}
\bibinfo{author}{Martin \surnamestart Hyland\surnameend}
  (\bibinfo{year}{2004}): \emph{\bibinfo{title}{Abstract Interpretation of
  Proofs: Classical Propositional Calculus}}.
\newblock In \bibinfo{editor}{Jerzy \surnamestart Marcinkowski\surnameend} \&
  \bibinfo{editor}{Andrzej \surnamestart Tarlecki\surnameend}, editors: {\sl
  \bibinfo{booktitle}{Computer Science Logic, 18th International Workshop,
  {CSL} 2004, Proceedings}}, {\sl \bibinfo{series}{Lecture Notes in Computer
  Science}} \bibinfo{volume}{3210}, \bibinfo{publisher}{Springer}, pp.
  \bibinfo{pages}{6--21}, \doi{10.1007/978-3-540-30124-0\_2}.
\newblock \urlprefix\url{https://doi.org/10.1007/978-3-540-30124-0\_2}.

\bibitemdeclare{article}{KellyLaplaza1980}
\bibitem{KellyLaplaza1980}
\bibinfo{author}{G.~M. \surnamestart Kelly\surnameend} \&
  \bibinfo{author}{M.~L. \surnamestart Laplaza\surnameend}
  (\bibinfo{year}{1980}): \emph{\bibinfo{title}{Coherence for compact closed
  categories}}.
\newblock {\sl \bibinfo{journal}{J. Pure Appl. Algebra}} \bibinfo{volume}{19},
  pp. \bibinfo{pages}{193--213}, \doi{10.1016/0022-4049(80)90101-2}.

\bibitemdeclare{book}{kock_2003}
\bibitem{kock_2003}
\bibinfo{author}{Joachim \surnamestart Kock\surnameend} (\bibinfo{year}{2003}):
  \emph{\bibinfo{title}{Frobenius Algebras and 2-D Topological Quantum Field
  Theories}}.
\newblock \bibinfo{series}{London Mathematical Society Student Texts},
  \bibinfo{publisher}{Cambridge University Press},
  \doi{10.1017/CBO9780511615443}.

\bibitemdeclare{incollection}{KrumlPaseka2008}
\bibitem{KrumlPaseka2008}
\bibinfo{author}{David \surnamestart Kruml\surnameend} \& \bibinfo{author}{Jan
  \surnamestart Paseka\surnameend} (\bibinfo{year}{2008}):
  \emph{\bibinfo{title}{Algebraic and categorical aspects of quantales}}.
\newblock In: {\sl \bibinfo{booktitle}{Handbook of algebra. {V}ol. 5}}, {\sl
  \bibinfo{series}{Handb. Algebr.}}~\bibinfo{volume}{5},
  \bibinfo{publisher}{Elsevier/North-Holland, Amsterdam}, pp.
  \bibinfo{pages}{323--362}, \doi{10.1016/S1570-7954(07)05006-1}.
\newblock \urlprefix\url{https://doi.org/10.1016/S1570-7954(07)05006-1}.

\bibitemdeclare{unpublished}{deLacroixSantocanale2022}
\bibitem{deLacroixSantocanale2022}
\bibinfo{author}{Cédric \surnamestart de~Lacroix\surnameend} \&
  \bibinfo{author}{Luigi \surnamestart Santocanale\surnameend}
  (\bibinfo{year}{2022}): \emph{\bibinfo{title}{Unitless {F}robenius
  quantales}}.
\newblock \bibinfo{note}{Preprint, available at
  \url{https://arxiv.org/pdf/2205.04111.pdf}}.

\bibitemdeclare{book}{maclane}
\bibitem{maclane}
\bibinfo{author}{Saunders \surnamestart Maclane\surnameend}
  (\bibinfo{year}{1978}): \emph{\bibinfo{title}{Categories for the working
  mathematician}}.
\newblock \bibinfo{series}{Graduate Texts in Mathematics},
  \bibinfo{publisher}{Springer-Verlag New York}.

\bibitemdeclare{book}{Maddux2006}
\bibitem{Maddux2006}
\bibinfo{author}{Roger~D. \surnamestart Maddux\surnameend}
  (\bibinfo{year}{2006}): \emph{\bibinfo{title}{Relation algebras}}.
\newblock {\sl \bibinfo{series}{Studies in Logic and the Foundations of
  Mathematics}} \bibinfo{volume}{150}, \bibinfo{publisher}{Elsevier B. V.,
  Amsterdam}.

\bibitemdeclare{article}{Raney60}
\bibitem{Raney60}
\bibinfo{author}{George~N. \surnamestart Raney\surnameend}
  (\bibinfo{year}{1960}): \emph{\bibinfo{title}{Tight {G}alois connections and
  complete distributivity}}.
\newblock {\sl \bibinfo{journal}{Trans. Amer. Math. Soc.}}
  \bibinfo{volume}{97}, pp. \bibinfo{pages}{418--426}, \doi{10.2307/1993380}.

\bibitemdeclare{article}{Resende2001}
\bibitem{Resende2001}
\bibinfo{author}{Pedro \surnamestart Resende\surnameend}
  (\bibinfo{year}{2001}): \emph{\bibinfo{title}{Quantales, finite observations
  and strong bisimulation}}.
\newblock {\sl \bibinfo{journal}{Theor. Comput. Sci.}}
  \bibinfo{volume}{254}(\bibinfo{number}{1-2}), pp. \bibinfo{pages}{95--149},
  \doi{10.1016/S0304-3975(99)00123-1}.
\newblock \urlprefix\url{https://doi.org/10.1016/S0304-3975(99)00123-1}.

\bibitemdeclare{book}{Rosenthal1990a}
\bibitem{Rosenthal1990a}
\bibinfo{author}{Kimmo~I. \surnamestart Rosenthal\surnameend}
  (\bibinfo{year}{1990}): \emph{\bibinfo{title}{Quantales and their
  applications}}.
\newblock {\sl \bibinfo{series}{Pitman Research Notes in Mathematics Series}}
  \bibinfo{volume}{234}, \bibinfo{publisher}{Longman Scientific \& Technical,
  Harlow}.

\bibitemdeclare{article}{Rowe1988}
\bibitem{Rowe1988}
\bibinfo{author}{K.~A. \surnamestart Rowe\surnameend} (\bibinfo{year}{1988}):
  \emph{\bibinfo{title}{Nuclearity}}.
\newblock {\sl \bibinfo{journal}{Canad. Math. Bull.}}
  \bibinfo{volume}{31}(\bibinfo{number}{2}), pp. \bibinfo{pages}{227--235},
  \doi{10.4153/CMB-1988-035-5}.

\bibitemdeclare{inproceedings}{SAN-2019-WORDS}
\bibitem{SAN-2019-WORDS}
\bibinfo{author}{Luigi \surnamestart Santocanale\surnameend}
  (\bibinfo{year}{2019}): \emph{\bibinfo{title}{On discrete idempotent paths}}.
\newblock In \bibinfo{editor}{Robert \surnamestart Mercaş\surnameend} \&
  \bibinfo{editor}{Daniel \surnamestart Reidenbach\surnameend}, editors: {\sl
  \bibinfo{booktitle}{{C}ombinatorics on {W}ords. {WORDS 2019}}}, {\sl
  \bibinfo{series}{Lecture Notes in Computer Science}} \bibinfo{volume}{11682},
  \bibinfo{publisher}{Springer, Cham}, pp. \bibinfo{pages}{312--325}.
\newblock \urlprefix\url{https://doi.org/10.1007/978-3-030-28796-2_25}.

\bibitemdeclare{inproceedings}{S_ACT2020}
\bibitem{S_ACT2020}
\bibinfo{author}{Luigi \surnamestart Santocanale\surnameend}
  (\bibinfo{year}{2020}): \emph{\bibinfo{title}{Dualizing sup-preserving
  endomaps of a complete lattice}}.
\newblock In \bibinfo{editor}{David~I. \surnamestart Spivak\surnameend} \&
  \bibinfo{editor}{Jamie \surnamestart Vicary\surnameend}, editors: {\sl
  \bibinfo{booktitle}{Proceedings of {ACT} 2020, Cambridge, USA, 6-10th July
  2020}}, {\sl \bibinfo{series}{{EPTCS}}} \bibinfo{volume}{333}, pp.
  \bibinfo{pages}{335--346}, \doi{10.4204/EPTCS.333.23}.
\newblock \urlprefix\url{https://doi.org/10.4204/EPTCS.333.23}.

\bibitemdeclare{inproceedings}{S_RAMICS2020}
\bibitem{S_RAMICS2020}
\bibinfo{author}{Luigi \surnamestart Santocanale\surnameend}
  (\bibinfo{year}{2020}): \emph{\bibinfo{title}{The Involutive Quantaloid of
  Completely Distributive Lattices}}.
\newblock In \bibinfo{editor}{Uli \surnamestart Fahrenberg\surnameend},
  \bibinfo{editor}{Peter \surnamestart Jipsen\surnameend} \&
  \bibinfo{editor}{Michael \surnamestart Winter\surnameend}, editors: {\sl
  \bibinfo{booktitle}{Proceedings of RAMiCS 2020, Palaiseau, France, April
  8-11, 2020 [postponed]}}, {\sl \bibinfo{series}{Lecture Notes in Computer
  Science}} \bibinfo{volume}{12062}, \bibinfo{publisher}{Springer}, pp.
  \bibinfo{pages}{286--301}, \doi{10.1007/978-3-030-43520-2\_18}.
\newblock \urlprefix\url{https://doi.org/10.1007/978-3-030-43520-2\_18}.

\bibitemdeclare{inproceedings}{S_RAMICS2021}
\bibitem{S_RAMICS2021}
\bibinfo{author}{Luigi \surnamestart Santocanale\surnameend}
  (\bibinfo{year}{2021}): \emph{\bibinfo{title}{Skew Metrics Valued in
  {S}ugihara Semigroups}}.
\newblock In \bibinfo{editor}{Uli \surnamestart Fahrenberg\surnameend},
  \bibinfo{editor}{Mai \surnamestart Gehrke\surnameend}, \bibinfo{editor}{Luigi
  \surnamestart Santocanale\surnameend} \& \bibinfo{editor}{Michael
  \surnamestart Winter\surnameend}, editors: {\sl
  \bibinfo{booktitle}{Relational and Algebraic Methods in Computer Science -
  19th International Conference, RAMiCS 2021, Marseille, France, November 2-5,
  2021, Proceedings}}, {\sl \bibinfo{series}{Lecture Notes in Computer
  Science}} \bibinfo{volume}{13027}, \bibinfo{publisher}{Springer}, pp.
  \bibinfo{pages}{396--412}, \doi{10.1007/978-3-030-88701-8\_24}.
\newblock \urlprefix\url{https://doi.org/10.1007/978-3-030-88701-8\_24}.

\bibitemdeclare{article}{DePaivaSchalk2004}
\bibitem{DePaivaSchalk2004}
\bibinfo{author}{Andrea \surnamestart Schalk\surnameend} \&
  \bibinfo{author}{Valeria \surnamestart de~Paiva\surnameend}
  (\bibinfo{year}{2004}): \emph{\bibinfo{title}{Poset-valued sets or how to
  build models for linear logics}}.
\newblock {\sl \bibinfo{journal}{Theor. Comput. Sci.}}
  \bibinfo{volume}{315}(\bibinfo{number}{1}), pp. \bibinfo{pages}{83--107},
  \doi{10.1016/j.tcs.2003.11.014}.
\newblock \urlprefix\url{https://doi.org/10.1016/j.tcs.2003.11.014}.

\bibitemdeclare{article}{Street2004}
\bibitem{Street2004}
\bibinfo{author}{Ross \surnamestart Street\surnameend} (\bibinfo{year}{2004}):
  \emph{\bibinfo{title}{Frobenius monads and pseudomonoids}}.
\newblock {\sl \bibinfo{journal}{J. Math. Phys.}}
  \bibinfo{volume}{45}(\bibinfo{number}{10}), pp. \bibinfo{pages}{3930--3948},
  \doi{10.1063/1.1788852}.
\newblock \urlprefix\url{https://doi.org/10.1063/1.1788852}.

\bibitemdeclare{article}{Yetter1990}
\bibitem{Yetter1990}
\bibinfo{author}{David~N. \surnamestart Yetter\surnameend}
  (\bibinfo{year}{1990}): \emph{\bibinfo{title}{Quantales and (Noncommutative)
  Linear Logic}}.
\newblock {\sl \bibinfo{journal}{J. Symb. Log.}}
  \bibinfo{volume}{55}(\bibinfo{number}{1}), pp. \bibinfo{pages}{41--64},
  \doi{10.2307/2274953}.
\newblock \urlprefix\url{https://doi.org/10.2307/2274953}.

\end{thebibliography}
